\documentclass[sigconf,natbib,pbalance]{acmart}
\usepackage[inline]{enumitem}
\usepackage{float}
\usepackage{stfloats}
\usepackage{latexsym}
\usepackage{listings}
\usepackage{longtable}
\usepackage{multicol}
\usepackage{url,tcolorbox}
\usepackage{graphicx}
\usepackage{wrapfig}
\usepackage{natbib}
\usepackage{booktabs,fvextra}
\usepackage{comment}
\usepackage{subcaption}
\usepackage{verbatim}
\usepackage{longtable}
\usepackage{tikz}
\usepackage{array}
\usepackage{enumitem,verbatimbox}
\usepackage{xspace}
\usepackage{algorithm} 
\usepackage{algpseudocode}
\usepackage{multirow}
\usepackage[table,xcdraw]{xcolor}
\usepackage{amsmath}
\usepackage{cleveref}
\usepackage[normalem]{ulem}

\AtBeginDocument{%
  }

\copyrightyear{2026}
\acmYear{2026}
\setcopyright{cc}
\setcctype{by}
\acmConference[ICTIR '26]{Proceedings of the 2026 International ACM SIGIR Conference on Innovative Concepts and Theories in Information Retrieval (ICTIR)}{July 25, 2026}{Melbourne, VIC, Australia}
\acmBooktitle{Proceedings of the 2026 International ACM SIGIR Conference on Innovative Concepts and Theories in Information Retrieval (ICTIR) (ICTIR '26), July 25, 2026, Melbourne, VIC, Australia}
\acmDOI{10.1145/3805713.3820437}
\acmISBN{979-8-4007-2600-2/2026/07}

\newcommand{\ie}{\emph{i.e.}}
\newcommand{\eg}{\emph{e.g.}}

\newcommand{\vs}{\emph{vs. }}
\newcommand{\incl}{\emph{incl. }}

\newenvironment{CRUPDATES}
{\comment}
{\endcomment}

\begin{document}

\title{From Noise to Order: Learning to Rank via Denoising Diffusion}

\author{Sajad Ebrahimi}
\orcid{0009-0003-1630-3938}
\affiliation{%
  \institution{University of Guelph}
  \city{Guelph}
  \state{ON}
  \country{Canada}
}

\author{Bhaskar Mitra}
\orcid{0000-0002-5270-5550}
\affiliation{%
  \institution{Independent Researcher}
  \city{Tiohtià:ke / Montréal}
  \state{QC}
  \country{Canada}
}

\author{Negar Arabzadeh}
\orcid{0000-0002-4411-7089}
\affiliation{%
  \institution{University of California, Berkeley}
  \city{Berkeley}
  \state{CA}
  \country{United States}
}

\author{Ye Yuan}
\orcid{0009-0001-3288-247X}
\affiliation{%
  \institution{McGill University, Mila}
  \city{Montréal}
  \state{QC}
  \country{Canada}
}

\author{Haolun Wu}
\orcid{0000-0001-6255-1535}
\affiliation{%
  \institution{Stanford University}
  \city{Stanford}
  \state{CA}
  \country{United States}
}

\author{Fattane Zarrinkalam}
\orcid{0000-0002-2102-9190}
\affiliation{%
  \institution{University of Guelph}
  \city{Guelph}
  \state{ON}
  \country{Canada}
}

\author{Ebrahim Bagheri}
\orcid{0000-0002-5148-6237}
\affiliation{%
  \institution{University of Toronto}
  \city{Toronto}
  \state{ON}
  \country{Canada}
}

\renewcommand{\shortauthors}{Sajad Ebrahimi et al.}

\begin{abstract}
In information retrieval (IR), learning-to-rank (LTR) methods have traditionally limited themselves to discriminative machine learning approaches that model the probability of the document being relevant to the query given some feature representation of the query-document pair.
In this work, we propose an alternative denoising diffusion-based deep generative approach to LTR that instead models the full joint distribution over feature vectors and relevance labels.
While in the discriminative setting, an over-parameterized ranking model may find different ways to fit the training data, we hypothesize that candidate solutions that can explain the full data distribution under the generative setting are better equipped to estimate relevance.
With this motivation, we propose DiffusionRank that extends TabDiff, an existing denoising diffusion-based generative model for tabular datasets, to create generative equivalents of classical discriminative pointwise and pairwise LTR objectives.
We conduct thorough empirical evaluation on four standard LTR datasets to demonstrate improvements from DiffusionRank models over their discriminative counterparts.
Our work points to a rich space for future research exploration on how we can leverage ongoing advancements in deep generative modeling approaches, such as diffusion, for LTR. 
We made our code publicly available at \url{https://github.com/sadjadeb/DiffusionRank}.
\vspace{-0.5em}
\end{abstract}


\vspace{-2em}

\maketitle

\section{Introduction}
\label{sec:intro}
In Information Retrieval (IR), Learning-to-Rank (LTR)~\citep{liu2009learning} is the task of constructing a model to estimate the relevance of an item (\eg, a document) with respect to an information need (\eg, expressed as a query by the user) so that the IR system can sort the items by their estimated relevance scores for presentation to the user.
These ranking models are typically trained on labeled datasets containing a set of query-document pairs and corresponding ground-truth relevance labels.
Traditionally, the ranking model is trained discriminatively to predict a real-valued score given the feature-vector.
Various objectives for training ranking models have been explored for LTR that \citet{liu2009learning} broadly categorize under:
\begin{enumerate*}[label=(\roman*)]
    \item pointwise,
    \item pairwise, and
    \item listwise loss functions.
\end{enumerate*}
Under these three categories, different loss functions and machine learning (ML) approaches---including support vector machines~\citep{yue2007support}, neural networks~\citep{burges2005learning}, and boosted decision trees~\citep{wu2010adapting}---have been explored. 
However, these explorations have historically been limited to discriminative ML approaches, where the ranker models the probability of the relevance label conditioned on the feature vector.

In this work, we propose an alternative deep generative approach to LTR that instead models the joint distribution over feature vectors and relevance labels.
Deep generative LTR opens up new avenues to explore how we can leverage recent advancements in deep generative modeling approaches, such as diffusion~\citep{sohl2015deep, song2019generative, ho2020denoising}, for ranking in IR.
Unlike discriminative training, where the model is trained only to predict relevance labels conditioned on the features, generative training requires the model to learn the full underlying data distribution, including the conditional distribution of subsets of features given other features (and optionally the label).
In the discriminative setting, an over-parameterized ranking model may find different ways to fit to the training data.
In the presence of a choice between these alternative solutions, we hypothesize that the solution that fits the full joint distribution under the generative setting leads to more accurate relevance estimation.

Generative modeling of different data modalities---\incl text~\citep{minaee2024large}, images~\citep{li2025comprehensive}, video~\citep{xing2024survey}, speech~\citep{cui2025recent}, and tabular data~\citep{wang2024challenges}---have recently demonstrated significant improvements in downstream applications.
Generative modeling of tabular datasets has found applications in missing value imputation~\citep{zheng2022diffusion}, training data augmentation~\citep{fonseca2023tabular}, and data privacy protection~\citep{hernandez2022synthetic, assefa2020generating}.
Several deep generative models have been proposed for modeling tabular datasets with autoregressive models~\citep{borisov2022language}, Variational Autoencoders (VAEs)~\citep{liu2023goggle}, Generative Adversarial Networks (GANs)~\citep{xu2019modeling}, and diffusion models~\citep{kotelnikov2023tabddpm, lee2023codi, kitouni2023disk, kim2022sos, zhang2024mixed, zheng2022diffusion, shi2025tabdiff}.
In this work, we extend generative approaches to modeling tabular data to LTR datasets to jointly model numerical LTR features and categorical relevance labels.
Specifically, we build on TabDiff's~\citep{shi2025tabdiff} continuous-time diffusion process to propose a family of diffusion-based LTR models that we call \textit{DiffusionRank}.

Tabular data presents unique challenges for generative modeling because they comprise heterogeneous columns with varied distributions and data types---\eg, in the LTR datasets, numerical features are continuous while categorical relevance labels are discrete.
TabDiff~\citep{shi2025tabdiff} employs a joint diffusion process to simultaneously perturb continuous and discrete data, and learns a joint denoising model for all modalities.
We extend TabDiff to LTR datasets where we consider the ranking features as continuous and the relevance label---which maybe either pointwise or a preference label between a pair of documents for a query---as discrete data.
This formulation, that we refer to as DiffusionRank, allows us to flexibly extend TabDiff's mixed-type diffusion process to develop generative equivalents to traditional discriminative LTR objectives.

\begin{CRUPDATES}
Furthermore, TabDiff employs a masked diffusion process for categorical data that involves a single-step denoising (\ie, \textit{unmasking}) at inference time.
This has important efficiency implications for DiffusionRank as we need to run the inference of the model only once to estimate the query-document relevance score, identical to their discriminative LTR counterparts.
\end{CRUPDATES}

Unlike TabDiff, which employs Transformers, we parameterize DiffusionRank with a feedforward network with a negligible increase in learnable parameters over an equivalent discriminative model, further ensuring that inference time costs are comparable for generative and discriminative LTR.
Furthermore, DiffusionRank does not mandate any specific base model and can flexibly incorporate new advances in neural architecture design.

To summarize our contributions:
We propose DiffusionRank, a diffusion-based deep generative LTR model, that scouts a potential path forward to leverage the emerging advancements in diffusion modeling for the LTR task.
Our contributions include the formalization of DiffusionRank and extensive empirical evaluation on four standard LTR datasets.
\begin{CRUPDATES}
to demonstrate improved accuracy of relevance estimation from generative LTR approaches.
\end{CRUPDATES}
In this preliminary work, we focus on the pointwise and pairwise settings to rigorously test the hypothesis that generative LTR approaches improve over their discriminative counterparts, and we expect future work to extend these methods to listwise and other advanced LTR settings to achieve state-of-the-art performances.
We believe that this work lays a foundation towards developing more expansive generative research agendas in IR.

\begin{CRUPDATES}
Next, we formally introduce some of the foundational concepts in LTR and diffusion processes that we build on in our current work, and discuss related literature, in Section~\ref{sec:prelim}.
Then, we describe the DiffusionRank model in Section~\ref{sec:model}.
In Section~\ref{sec:experiments}, we describe our experimental methodology and evaluation protocols, before presenting our results and analysis in Section~\ref{sec:results}.
Finally, we conclude in Section~\ref{sec:conclusion} with a discussion on potential new research directions for future work on generative approaches to ranking in IR.
\end{CRUPDATES}
\vspace{-0.5em}
\section{Preliminaries and Related Work}
\label{sec:prelim}
In order to situate our work within the broader landscape of generative approaches to learning-to-rank, we first establish the necessary background for our proposed approach; this section therefore both introduces the key preliminaries we build on and clarifies how our method relates to—and differs from—prior work.

\subsection{Learning-to-Rank}
\label{sec:related-ltr}
Formally, let a rankable document $d$ in context of the query $q$ be represented by a feature vector $\vec{x}_{q,d} \in \mathbb{R}^n$, and its ground-truth relevance label with respect to $q$ be represented as $y_{q,d} \in \{0, 1, \ldots, R-1\}$, assuming a $R$-point labeling scheme.
The feature vector representing a query-document pair may comprise of manually-designed numeric features---\eg, BM25~\citep{robertson2009probabilistic} and PageRank~\citep{brin1998anatomy}---or, one-hot encoding of query and document tokens, in representation-learning ranking models~\citep{mitra2018introduction}.
The labeling scheme, in its simplified form, may use binary relevance assessments, \ie, $R=2$ and $y_{q,d} \in \{0, 1\}$, categorizing each document to be either \textit{non-relevant} or \textit{relevant} to the query.
Or alternatively, it can employ graded-relevance assessments---\eg, the $5$-point labeling scheme from Bing~\citep{mcgee2012yes} that categorizes each document's relevance to the query as one of $\{$Perfect, Excellent, Good, Fair, Bad$\}$.
The ranking model $f: \vec{x}_{q,d} \to \mathbb{R}$ takes the feature-vector $\vec{x}_{q,d}$ as input and predicts a real-valued score $s_{q,d} \in \mathbb{R}$ commensurate with its estimate of the document's relevance to the query.
The ranking model may be trained with pointwise, pairwise, or listwise objectives~\citep{liu2009learning, mitra2018introduction}.

\subsubsection{Pointwise objectives}
\label{sec:related-ltr-pointwise}
In pointwise training, the ranking model is optimized to predict the ground-truth label for a query-document pair.
If the relevance label is categorical, then its prediction can be treated as a multiclass classification problem~\citep{li2007mcrank}.
The model under this setting estimates the probability distribution over label categories and can be trained using the cross-entropy (CE) loss.
\begin{align}
    \mathcal{L}_\text{pointwise-CE} &= -\text{log}\big(p(y_{q,d} | \vec{x}_{q,d})\big)
\end{align}
If the relevance labels are binary, then the predicted probability of the document being relevant to the query can be directly used for ranking.
However, for graded-relevance labeling schemes, the probability distribution over the label categories must be aggregated to generate a single ranking score.

Pointwise ranking models can also be trained using regression objectives, such as the squared loss, where $y_{q,d}$ and $s_{q,d}$ are either represented as absolute values~\citep{cossock2006subset} or as one-hot encodings~\citep{fuhr1989optimum}.
\begin{align}
    \mathcal{L}_\text{pointwise-squared} &= \|y_{q,d} - s_{q,d}\|^2
\end{align}

\subsubsection{Pairwise objectives}
\label{sec:related-ltr-pairwise}
In contrast, pairwise training~\citep{chen2009ranking, herbrich2000large, freund2003efficient, burges2005learning} optimizes the ranking model to minimize the number of preference errors between pairs of documents for the same query (\ie, when $y_{q,d_i} > y_{q,d_j}$ but $s_{q,d_i} < s_{q,d_j}$) in the training data.
Pairwise ranking losses generally take the following form`\citep{chen2009ranking}.
\begin{align}
    \mathcal{L}_\text{pairwise} &= \phi(s_{q,d_i} - s_{q,d_j})
\end{align}
Where, $\phi$ can be the Hinge function~\citep{herbrich2000large}, the exponential function~\citep{freund2003efficient}, or the logistic function~\citep{burges2005learning}.
When the logistic function is used, the resulting loss function is referred to as RankNet~\cite{burges2005learning}.
If we sort the pair of documents, such that $y_{q,d_i} > y_{q,d_j}$, then
\begin{align}
    \mathcal{L}_\text{RankNet} &= -\text{log}\big(1 + e^{-(s_{q,d_i} - s_{q,d_j}) / \tau}\big)
\end{align}

\bigskip\noindent
In Section~\ref{sec:model}, we define diffusion-based generative LTR objectives corresponding to the discriminative losses $\mathcal{L}_\text{pointwise-CE}$ and $\mathcal{L}_\text{RankNet}$.

\begin{CRUPDATES}
\bigskip\noindent
\textit{\textbf{Specialized LTR. }}
There is also an extensive body of work specializing LTR to different settings.
This includes LTR trained with different sources of supervision (\eg, online interactions with users~\citep{hofmann2013fast} and biased feedback data~\citep{joachims2017unbiased}), LTR with specific attributes (\eg, stochasticity~\citep{diaz2020evaluating, bruch2020stochastic}, uncertainty-awareness~\citep{cohen2021not}, and interpretability~\citep{zhuang2021interpretable}), and LTR with additional constraints (\eg, diversity~\citep{radlinski2008learning} and fairness~\citep{singh:fair-pg-rank, diaz2020evaluating}).
These topics are out of scope for our current discussion but the application of generative LTR to these specialized settings may offer interesting future directions for research.
\end{CRUPDATES}

\begin{figure*}
    \centering
    \includegraphics[width=0.75\linewidth]{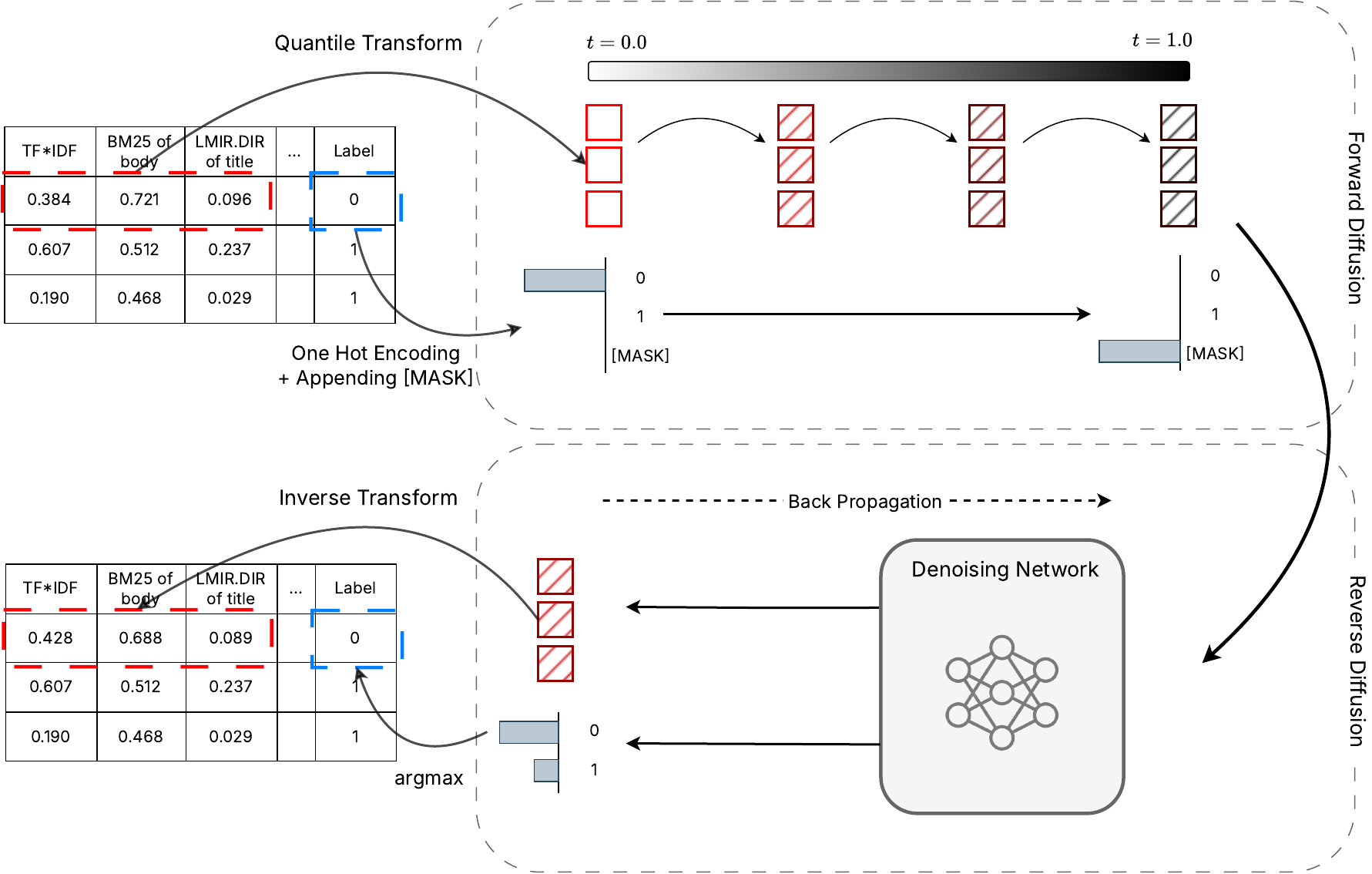}
    \caption{Overview of DiffusionRank (pointwise). We model LTR as denoising diffusion process over tabular feature-label tuples: a forward process progressively corrupts the input (Gaussian noise for numerical features and masking for categorical variables, including relevance labels), and a learned reverse process denoises to recover clean samples.}

    \label{fig:diffusionrank_pipeline}
\end{figure*}

\subsection{Diffusion modeling}
\label{sec:related-diffusion}

\subsubsection{Denoising Diffusion Probabilistic Models}
\label{sec:related-diffusion-ddpm}

Denoising Diffusion Probabilistic Models (DDPMs)~\citep{sohl2015deep, ho2020denoising} are a class of latent variable generative models inspired by non-equilibrium thermodynamics.
Unlike GANs that learn an implicit generative function, diffusion models are likelihood-based and learn to reverse a gradual noise-adding process.
The framework consists of two processes: a forward diffusion process and a reverse denoising process.

\paragraph{Forward Process}
The forward process is a fixed Markov chain that gradually adds Gaussian noise to the data $\vec{x}_0 \sim q(\vec{x}_0)$ over $T$ steps.
At each step $t$, the transition is defined as:
\begin{align}
    q(\vec{x}_t | \vec{x}_{t-1}) &= \mathcal{N}(\vec{x}_t; \sqrt{1 - \beta_t}\vec{x}_{t-1}, \beta_t \mathbf{I})
\end{align}
where $\{\beta_t\}_{t=1}^T$ is a predefined variance schedule.
Crucially, this process allows us to sample $\vec{x}_t$ at any arbitrary time step $t$ directly from $\vec{x}_0$ in closed form:
\begin{align}
    q(\vec{x}_t | \vec{x}_0) &= \mathcal{N}(\vec{x}_t; \sqrt{\bar{\alpha}_t}\vec{x}_0, (1 - \bar{\alpha}_t)\mathbf{I})
\end{align}
where $\alpha_t = 1 - \beta_t$ and $\bar{\alpha}_t = \prod_{s=1}^t \alpha_s$. As $T \to \infty$, the data $\vec{x}_0$ is transformed into pure Gaussian noise $\vec{x}_T \sim \mathcal{N}(\mathbf{0}, \mathbf{I})$.

\paragraph{Reverse Process}
The goal of the generative model is to reverse this process, starting from pure noise $\vec{x}_T$ and sequentially denoising it to recover a sample from the data distribution.
Since the true posterior $q(\vec{x}_{t-1}|\vec{x}_t)$ is intractable, DDPMs learn a parameterized approximation $p_\theta(\vec{x}_{t-1} | \vec{x}_t)$:
\begin{align}
    p_\theta(\vec{x}_{t-1} | \vec{x}_t) &= \mathcal{N}(\vec{x}_{t-1}; \boldsymbol{\mu}_\theta(\vec{x}_t, t), \boldsymbol{\Sigma}_\theta(\vec{x}_t, t))
\end{align}
To train the model, \citet{ho2020denoising} simplified the variational lower bound on the negative log-likelihood.
Instead of predicting the mean $\boldsymbol{\mu}_\theta$ directly, the model $\boldsymbol{\epsilon}_\theta(\vec{x}_t, t)$ is typically trained to predict the noise $\boldsymbol{\epsilon}$ that was added to $\vec{x}_0$ to generate $\vec{x}_t$.
The simplified training objective is a mean-squared error loss:
\begin{align}
    \mathcal{L}_\text{simple} &= \mathbb{E}_{t, \vec{x}_0, \boldsymbol{\epsilon}} \big[ \| \boldsymbol{\epsilon} - \boldsymbol{\epsilon}_\theta(\vec{x}_t, t) \|^2 \big]
\end{align}
While originally formulated for continuous data using Gaussian noise, this framework has recently been extended to continuous-time stochastic differential equations (SDEs)~\citep{song2021scorebased} and other data types~\citep{yang2023diffusurvey}.
In this work, we leverage these advancements to model the joint distribution of ranking features and relevance labels.

\subsubsection{Modeling Tabular Data With TabDiff}
\label{sec:related-diffusion-tabdiff}
TabDiff~\citep{shi2025tabdiff} operates on mixed-tabular datasets containing numerical and categorical features.
Let $\vec{z} = [\vec{x}, \vec{y}]$ represent a data sample consisting of a concatenation of vectors of $C_\text{num}$ continuous numerical features $\vec{x}$ and $C_\text{cat}$ categorical features $\vec{y}$.
The vector of categorical features $\vec{y}$ is in turn are a concatenation of the $C_\text{cat}$ individual categorical features $\vec{y}_i \in \{0, 1\}^{(c_i+1)}$ which are one-hot encoded, with $c_i$ as the number of categories in feature $\vec{y}_i$.
The one-hot encoding of $\vec{y}_i$ includes an additional dimension to indicate the [MASK] state for features during the masked diffusion process.

TabDiff, like other diffusion models, is a likelihood-based generative model that learns the data distribution using forward and backward Markov processes.
TabDiff gradually corrupts the data in the forward process and learns a denoising model to recover the original data in the reverse process.
Let $\{\vec{z}_t: t \sim [0,1]\}$ denote a sequence of data in the diffusion process where $t \in [0, 1]$ is a continuous time variable, such that $\vec{z}_0 \sim p_0$ is a real data sample and $\vec{z}_1 \sim p_1$ is pure noise from a prior distribution.
We can represent the forward diffusion process as follows:
\begin{align}
    q(\vec{z}_t | \vec{z}_0) = q(\vec{x}_t | \vec{x}_0, \sigma_\text{num}) \cdot q(\vec{y}_t | \vec{y}_0, \sigma_\text{cat})
\end{align}
Where, $\sigma_\text{num}$ and $\sigma_\text{cat}$ are the diffusion schedules for numerical and categorical features, respectively.
The true reverse process can consequently be represented as:
\begin{align}
    q(\vec{z}_s | \vec{z}_t, \vec{z}_0) = q(\vec{x}_s | \vec{z}_t, \vec{z}_0) \cdot q(\vec{y}_s | \vec{z}_t, \vec{z}_0)
\end{align}
Where, $s$ and $t$ represent two arbitrary time steps such that $0 < s < t < 1$.
TabDiff learns a denoising model $p_\theta(\vec{z}_s | \vec{z}_t)$ to approximate the true posterior $q(\vec{z}_s | \vec{z}_t, \vec{z}_0)$.
This denoising model is learnt by minimizing the following loss:
\begin{align}
    \mathcal{L}_\text{TabDiff} &= \lambda_\text{num} \cdot \mathcal{L}_\text{num} + \lambda_\text{cat} \cdot \mathcal{L}_\text{cat}
\end{align}
Where $\mathcal{L}_\text{num}$ and $\mathcal{L}_\text{cat}$ represent the loss components corresponding to the numerical and categorical features, respectively, and $\lambda_\text{num}$ and $\lambda_\text{cat}$ are two weight terms.
Let, $\vec{\chi}$ and $\vec{\psi}$ denote the numerical and categorical part of the denoising model's output, respectively.
We next describe how we compute $\mathcal{L}_\text{num}$ and $\mathcal{L}_\text{cat}$ with respect to the denoising model outputs $\vec{\chi}$ and $\vec{\psi}$.

\paragraph{Gaussian diffusion for numerical features}
TabDiff's forward diffusion process gradually corrupts the numerical features by adding sampled noise to them, as represented below:
\begin{align}
    \vec{x}_t = \vec{x}_0 + \sigma_\text{num}(t) \cdot \vec{\epsilon}, \;\;\;\;\vec{\epsilon} \sim \mathcal{N}(\vec{0}, \vec{\mathrm{I}}_{C_\text{num}})
\end{align}
The numerical part of the denoising model then tries to predict the noise that was added, and trains by minimizing the following loss:
\begin{align}
    \mathcal{L}_\text{num} = \mathbb{E}_{t \sim U[0, 1]} \mathbb{E}_{(\vec{z}_t, \vec{z}_0) \sim q(\vec{z}_t, \vec{z}_0)} \|\vec{\chi}(t) - \vec{\epsilon}\|^2_2
\end{align}

\paragraph{Masked diffusion for categorical features}
For categorical features, the forward diffusion process is defined as a masking process, represented as follows:
\begin{align}
    q(\vec{x}_t|\vec{x}_0) = \text{Cat}\big(\vec{x}_t; \alpha_t \cdot \vec{x}_0 + (1 - \alpha_t)\mathrm{m}\big)
\end{align}
Where, $\text{Cat}\big(;\pi)$ is a categorical distribution over the classes with probabilities given by $\pi$, and $\alpha_t \in [0,1]$ is a strictly-decreasing function of $t$, with the additional constraints that $\alpha_0 \approx 1$ and $\alpha_1 \approx 0$.
At each diffusion step, the feature is corrupted by being changed to the [MASK] state with a probability of $(1 - \alpha_t)$ and then remains as such till $t=1$.
At time $t=0$, all categorical features are unmasked; and at $t=1$, all of them are masked.

In the reverse denoising process, the model aims to recover the original feature values from the masked state. When a feature is masked in the input, the model predicts it conditioned on the remaining noisy features at time $t$. During training, the loss is computed only on predictions corresponding to inputs in the masked state, ensuring that the model learns to unmask corrupted features rather than trivially copying unmasked ones.
The categorical part of the denoising model is trained by minimizing the following loss, where $\alpha'_t$ is the first order derivative of $\alpha_t$:
\begin{align}
    \mathcal{L}_\text{cat} = \mathbb{E}_{t \sim U[0, 1]} \mathbb{E}_{(\vec{z}_t, \vec{z}_0) \sim q(\vec{z}_t, \vec{z}_0)} \Big[ \frac{\alpha'_t}{1 - \alpha_t} \cdot \text{log} \langle\vec{\psi}(t), \vec{y}_0\rangle\Big]
\end{align}

\bigskip\noindent
We point the reader to \citet{shi2025tabdiff} for further details about TabDiff.

\subsection{Alternative Approach(es) to Generative LTR}
\label{sec:related-generative-ltr}
While the LTR research community has focused almost exclusively on discriminative ML approaches, IRGAN~\citet{wang2017irgan} and subsequent follow up works---\eg,~\citep{yu2023depth, park2019adversarial, li2022learning, deshpande2020evaluating, jain2020improving, lu2019psgan, yu2021diagnostic, li2024listwise}---constitute one strand of IR research that has explored Generative Adversarial Networks (GANs)~\citep{goodfellow2014generative} for ranking.
However, that line of work has focused more on the \textit{adversarial} aspects of GAN, rather than its \textit{generative} aspects.
The IRGAN formulation involves two models:
\begin{enumerate*}[label=(\alph*)]
    \item a generator model $p_\theta(d|q,y)$ which identifies non-relevant documents for a query from a candidate pool that closely resemble relevant documents in some feature space, and
    \item a discriminator model that tries to discriminate between the relevant and the adversarially selected non-relevant document.
\end{enumerate*}
Unlike our work, IRGAN does not learn a joint distribution over query-document features $\vec{x}_{q,d}$ and the relevance label $y_{q,d}$.
Therefore, further comparisons with IRGAN are out of scope for our work.

The other strand of related work includes the use of generative large language models (LLMs) for ranking~\citep{sun2023chatgpt, qin2024large, zhuang2024beyond, gao2025llm4rerank, wang2025realm}.
These approaches prompt LLMs trained with generative modeling objectives to estimate query-document relevance.
However, these approaches do not focus on \textit{learning} to rank, and hence are also out of scope for our current work.

\begin{figure}
    \centering
    \hspace{15pt}
    \begin{subfigure}[t]{0.45\linewidth}
        \includegraphics[height=4cm]{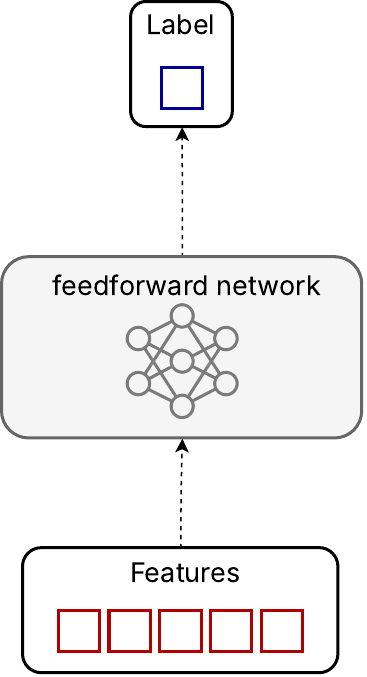}
        \caption{}
        \label{fig:diffusionrank_models_disc}
    \end{subfigure}
    \hspace{-15pt}
    \begin{subfigure}[t]{0.45\linewidth}
        \includegraphics[height=4cm]{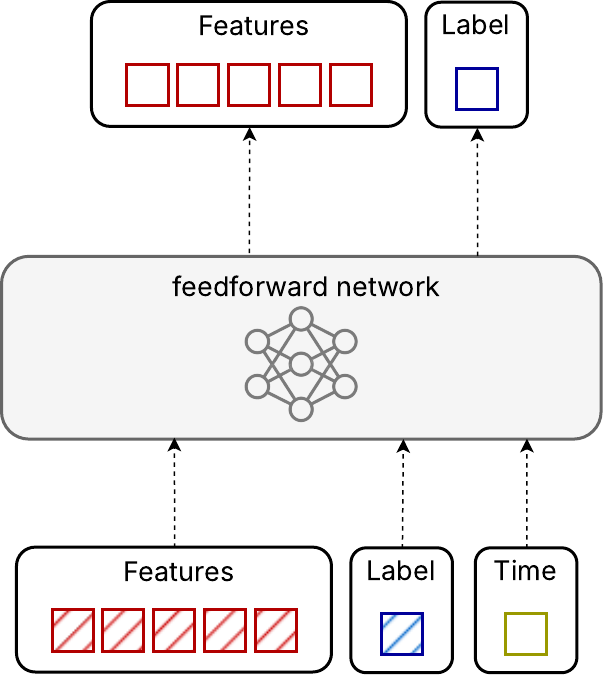}
        \caption{}
        \label{fig:diffusionrank_models_gen_point}
    \end{subfigure}
    \hfill
    \vspace{10pt}
    \begin{subfigure}[t]{0.9\linewidth}
        \includegraphics[width=\textwidth]{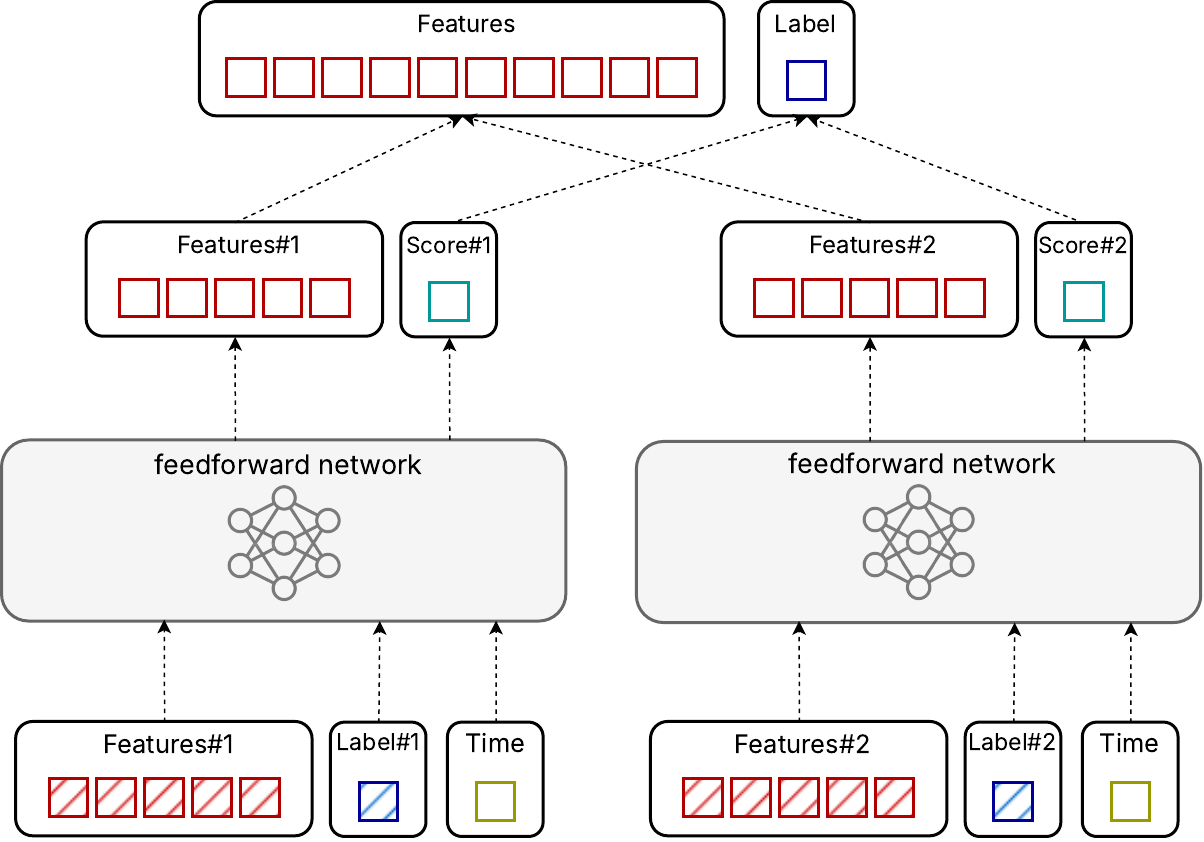}
        \caption{}
        \label{fig:diffusionrank_models_gen_pair}
    \end{subfigure}
    \caption{Architectural differences between discriminative and DiffusionRank models.
(a) Standard discriminative LTR model, which takes only query-document features as input and predicts a relevance score or label.
(b) Pointwise DiffusionRank, where the model additionally conditions on the (possibly masked) relevance label and diffusion time step, and jointly predicts the relevance label and the noise added to features.
(c) Pairwise DiffusionRank, which applies the pointwise denoising model independently to a document pair with tied label masking, producing scores for both documents while learning from noisy feature representations.}
    \label{fig:diffusionrank_models}
    \vspace{-1em}
\end{figure}

\section{DiffusionRank}
\label{sec:model}
DiffusionRank extends TabDiff to LTR datasets by treating the ranking features and the relevance labels (both pointwise or pairwise) as numerical and categorical tabular data, respectively.
\Cref{fig:diffusionrank_pipeline} shows the forward and reverse diffusion processes of our training pipeline.
Like TabDiff, we employ Gaussian diffusion over the LTR features and masked diffusion for predicting the relevance label.

We parameterize the diffusion process with a feedforward network, an architecture choice that is common in the LTR settings.
DiffusionRank can flexibly incorporate any alternative neural architectures as long as they can be extended to:
\begin{enumerate*}[label=(\roman*)]
    \item Accept the label, which may be masked or unmasked, and the time step $t$ of the diffusion process as additional inputs, and
    \item predict the noise over the features jointly with the relevance labels.
\end{enumerate*}
Figure~\ref{fig:diffusionrank_models} sketches out these necessary modifications to the base model from the discriminative LTR setting.
Next, we elaborate on how we instantiate this framework for pointwise and pairwise LTR settings.

\subsection{Pointwise DiffusionRank}
\label{sec:model-pointwise}
Pointwise DiffusionRank is a straightforward application of the framework just described.
The LTR model $f: \vec{x}_{q,d} \to \mathbb{R}$ from the discriminative setting is modified to operate as a denoising model $\bar{f}: [\vec{x}_{q,d}, y_{q,d}, t] \to [\mathbb{R}^{C_\text{num}}, \mathbb{R}^{C_\text{cat}+1}]$ as shown in Figure~\ref{fig:diffusionrank_models_gen_point}, where $C_\text{num}$ is the number of LTR features in the dataset and $C_\text{cat} = k$ for a $k$-point relevance labeling scheme.
The relevance label, as both input to and output of the model, is one-hot encoded with an extra [MASK] dimension.
As part of the diffusion process, the model is trained to predict the relevance label, given
\begin{enumerate*}[label=(\roman*)]
    \item the features corrupted by adding noise,
    \item time step $t$, and
    \item the label as masked in the input.
\end{enumerate*}
This is equivalent to training a discriminative model with the $\mathcal{L}_\text{pointwise-CE}$ loss, where the features have been corrupted and two additional features (the masked label and the time step $t$) are added as input.
Furthermore, the model also predicts the estimated noise with respect to the feature vector which encourages the model to learn the joint distribution of features and labels in the dataset.

At inference time, we provide as input
\begin{enumerate*}[label=(\roman*)]
    \item the uncorrupted features,
    \item time step $t=0$, and
    \item {[MASK]} as the label;
\end{enumerate*}
and use the model to directly predict the relevance of the query-document pair.
The part of the final layer of the model that predicts the feature noise can be safely ignored at inference time to keep the runtime cost almost exactly similar as that of its discriminative counterpart.
To further ensure comparable inference-time compute and time costs between the generative and discriminative settings, we do not use the backward stochastic sampler and the guidance classifier employed by \citet{shi2025tabdiff} in their original work.

\subsection{Pairwise DiffusionRank}
\label{sec:model-pairwise}
Unlike in the pointwise setting, in pairwise training we have two feature vectors $\vec{x}'_{q,d_i}$ and $\vec{x}'_{q,d_j}$ corresponding to the pair of documents for the same query, and a preference label $y$ denoting whether document $d_i$ is more relevant than $d_j$, or not.
From the TabDiff perspective, we consider $\vec{x}_t = [\vec{x}'_{q,d_i,t}, \vec{x}'_{q,d_j,t}]$ as the concatenation of the two feature vectors; and in the Gaussian diffusion process, the noise added to the two feature vectors are sampled independently.

In spite of the pairwise setting, we employ a pointwise model $\bar{f}: [\vec{x}_{q,d}, y_{q,d}, t] \to [\mathbb{R}^{C_\text{num}}, \mathbb{R}]$ for the denoising.
This is similar to standard discriminative pairwise settings, where the loss is pairwise (\ie, considers $\langle q, d_i, d_j \rangle$) but the models themselves are typically pointwise (\ie, they take $\langle q, d \rangle$ as input).
This pointwise model is applied to $\vec{x}'_{q,d_i}$ and $\vec{x}'_{q,d_j}$ independently in parallel.
The labels for the documents are tied-masked---\ie, they are either both masked or unmasked at the same time. 
When unmasking the label as part of the reverse diffusion process, the preference label prediction $\vec{\psi} = [s_{q,d_i}, s_{q,d_j}]$ is a concatenation of the scores $s_{q,d_i}$ and $s_{q,d_j}$.
This is similar to training the model with $\mathcal{L}_\text{RankNet}$ with the addition of noise to the features and the additional Gaussian diffusion loss as a component of the overall optimization objective.

\begin{table}[t]
\centering
\caption{Summary statistics of the LTR datasets used in our evaluation. All reported values correspond to Fold 1}
\label{tab:dataset-stats}
\resizebox{\columnwidth}{!}{
\begin{tabular}{lcccc}
\hline
                                & WEB30K & WEB10K & Istella-S & MQ2007 \\ \hline
\textbf{Queries (Train)}        & 18,919      & 6,000       & 19,245    & 1,017  \\
\textbf{Queries (Val)}          & 6,306       & 2,000       & 7,211     & 339    \\
\textbf{Queries (Test)}         & 6,306       & 2,000       & 6,562     & 336    \\ \hline
\textbf{Data Points (Total)}    & 3,771,125   & 1,200,192   & 3,408,630 & 69,623 \\
\textbf{Per Query (avg.)}       & 119.60      & 120.01      & 103.23    & 41.14  \\ \hline
\textbf{Features ($F$)}         & 136         & 136         & 220       & 46     \\
\textbf{Relevance Labels ($R$)} & 5           & 5           & 5         & 3      \\ \hline
\end{tabular}}
\vspace{-1em}
\end{table}

\bigskip\noindent
In this work, we designed diffusion-based counterparts to $\mathcal{L}_\text{pointwise-CE}$ and $\mathcal{L}_\text{RankNet}$ to demonstrate the generality of our method and its relevance to different LTR approaches.
We leave similar adaptation of other discriminative LTR objectives, including listwise losses, into the DiffusionRank framework for future work.

\section{Experiments}
\label{sec:experiments}

\subsection{Data}
\label{sec:experiments-data}
We evaluate our approach on four widely used LTR datasets: MQ2007 partition of LETOR 4.0, MSLR-WEB10K, MSLR-WEB30K~\cite{qin2013introducing}, and Istella-S~\cite{Lucchese2016PostLearningOO}. MQ2007, MSLR-WEB10K, and MSLR-WEB30K are constructed from real Bing search logs, with MSLR-WEB10K being approximately fourteen times larger than MQ2007, whereas Istella-S is derived from the Istella search engine and provides a large-scale dataset with over 3.4 million query-document pairs.

Each query-document pair in MQ2007 is represented by 46 numerical features, and relevance labels are graded from 0 to 2. In MSLR-WEBs, each pair has 136 features, and relevance labels range from 0 (irrelevant) to 4 (perfectly relevant). For Istella-S, each pair is represented by 220 features, and relevance labels follow the same 0 to 4 grading scale. Table~\ref{tab:dataset-stats} summarizes dataset statistics.

Our pointwise formulation assumes binary relevance labels in the training data to simplify the joint distribution modeled by the diffusion process, allowing the model to efficiently capture the fundamental distinction between relevant and non-relevant documents.
So, for pointwise training, we binarize the relevance labels in our datasets: for MQ2007, 0 $\to$ 0 and 1,2 $\to$ 1; for MSLR-WEBs and Istella-S, 0,1 $\to$ 0 and 2,3,4 $\to$ 1. The original labels were retained for evaluation. All features were transformed using a Quantile Transformer. The same transformation was also applied when training the baseline models to ensure a fair comparison.

For experiments analyzing the effect of training data size, we randomly sampled subsets of training data and used the same samples across all runs. This ensures all models were trained on exactly the same data, while the validation and test remained unchanged.

\begin{figure*}[t]
    \centering
    \begin{minipage}{0.7\textwidth}
        \centering
        \begin{subfigure}[t]{0.47\linewidth}
            \centering
            \includegraphics[width=\linewidth]{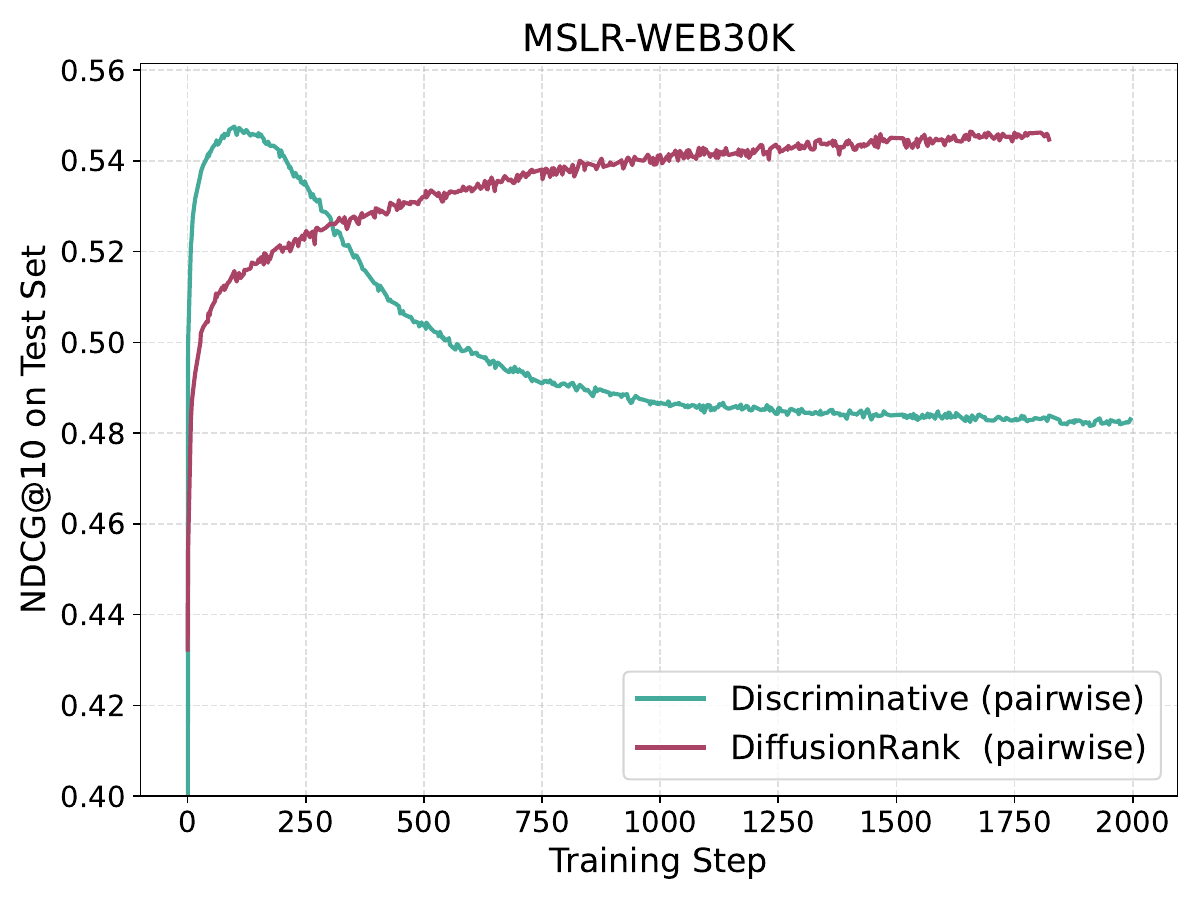}
            \caption{}
            \label{fig:training-mslr}
        \end{subfigure}
        \hfill
        \begin{subfigure}[t]{0.47\linewidth}
            \centering
            \includegraphics[width=\linewidth]{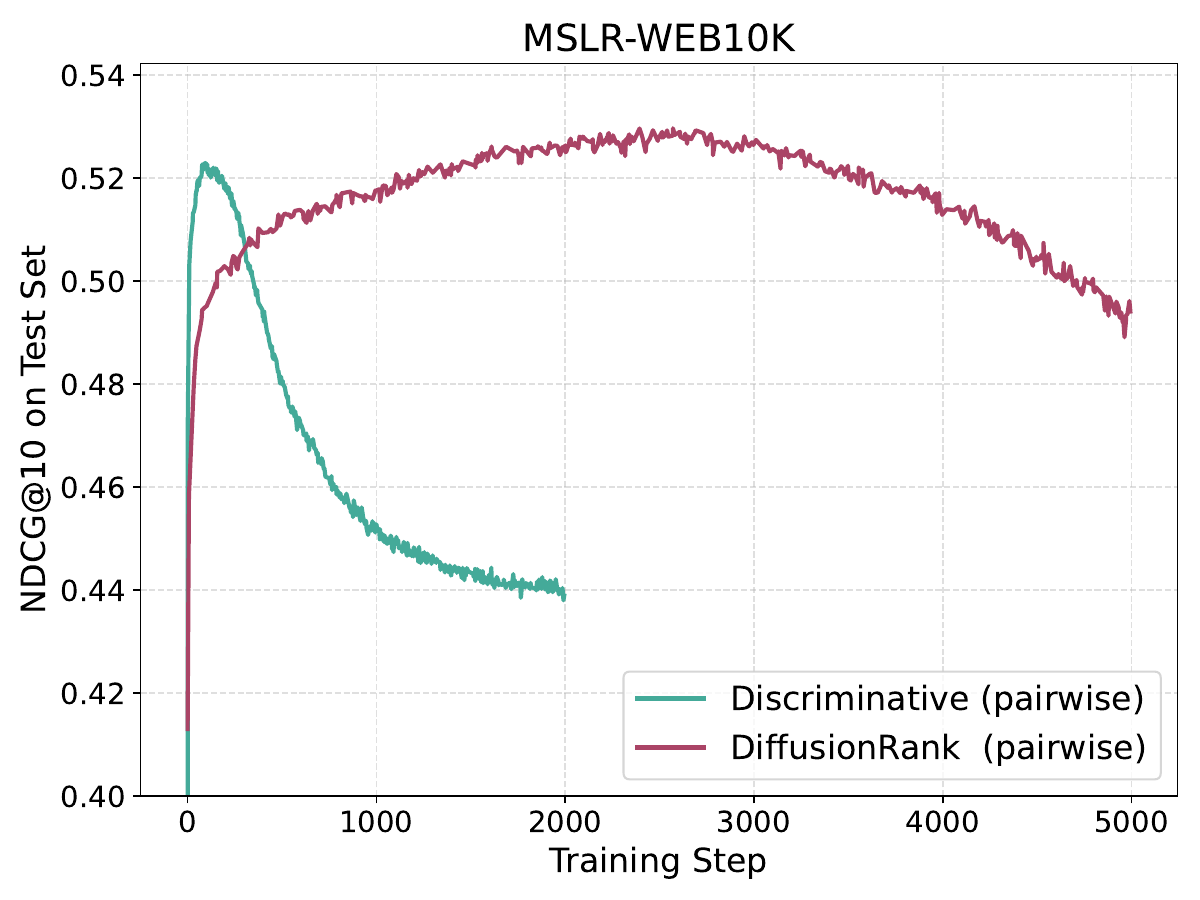}
            \caption{}
            \label{fig:training-mq2007}
        \end{subfigure}
        \hfill
        \begin{subfigure}[t]{0.47\linewidth}
            \centering
            \includegraphics[width=\linewidth]{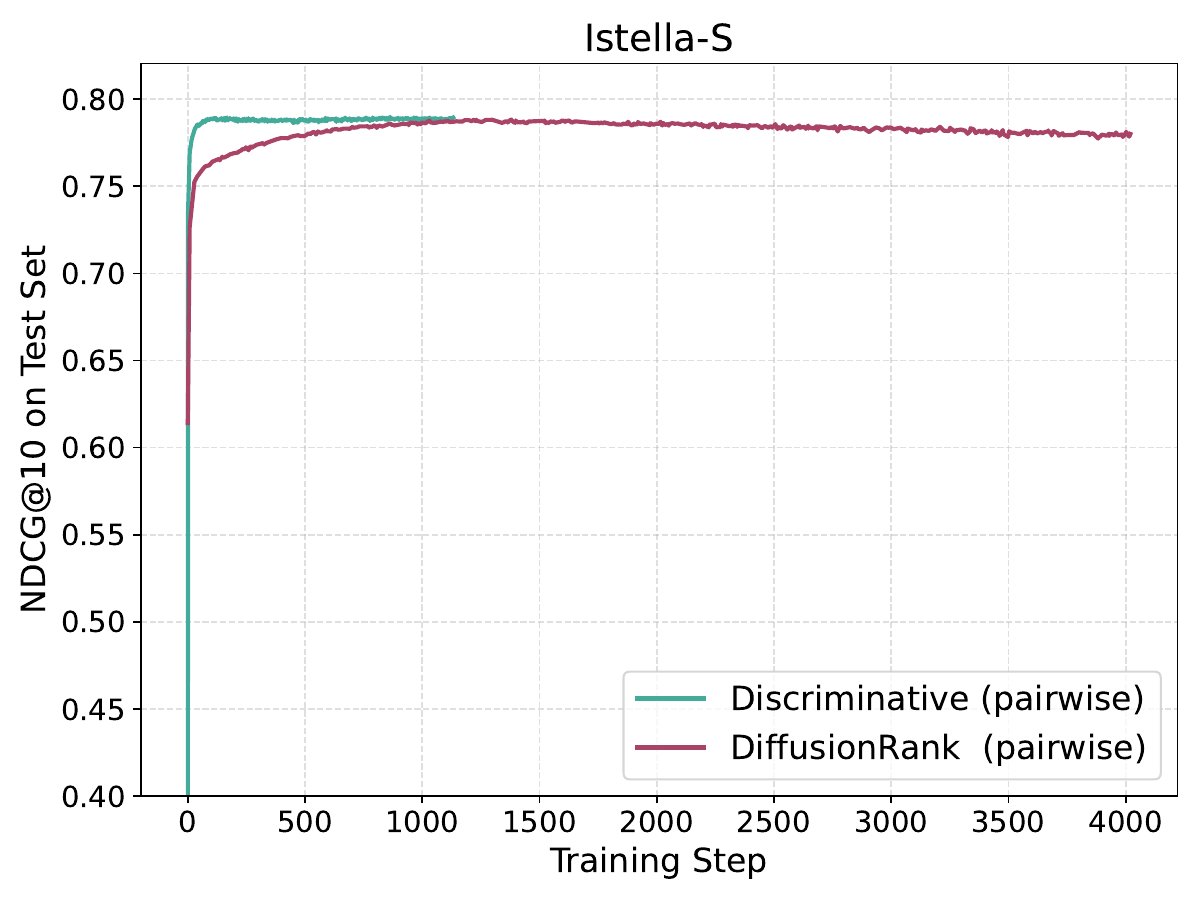}
            \caption{}
            \label{fig:training-mq2008}
        \end{subfigure}
        \hfill
        \begin{subfigure}[t]{0.47\linewidth}
            \centering
            \includegraphics[width=\linewidth]{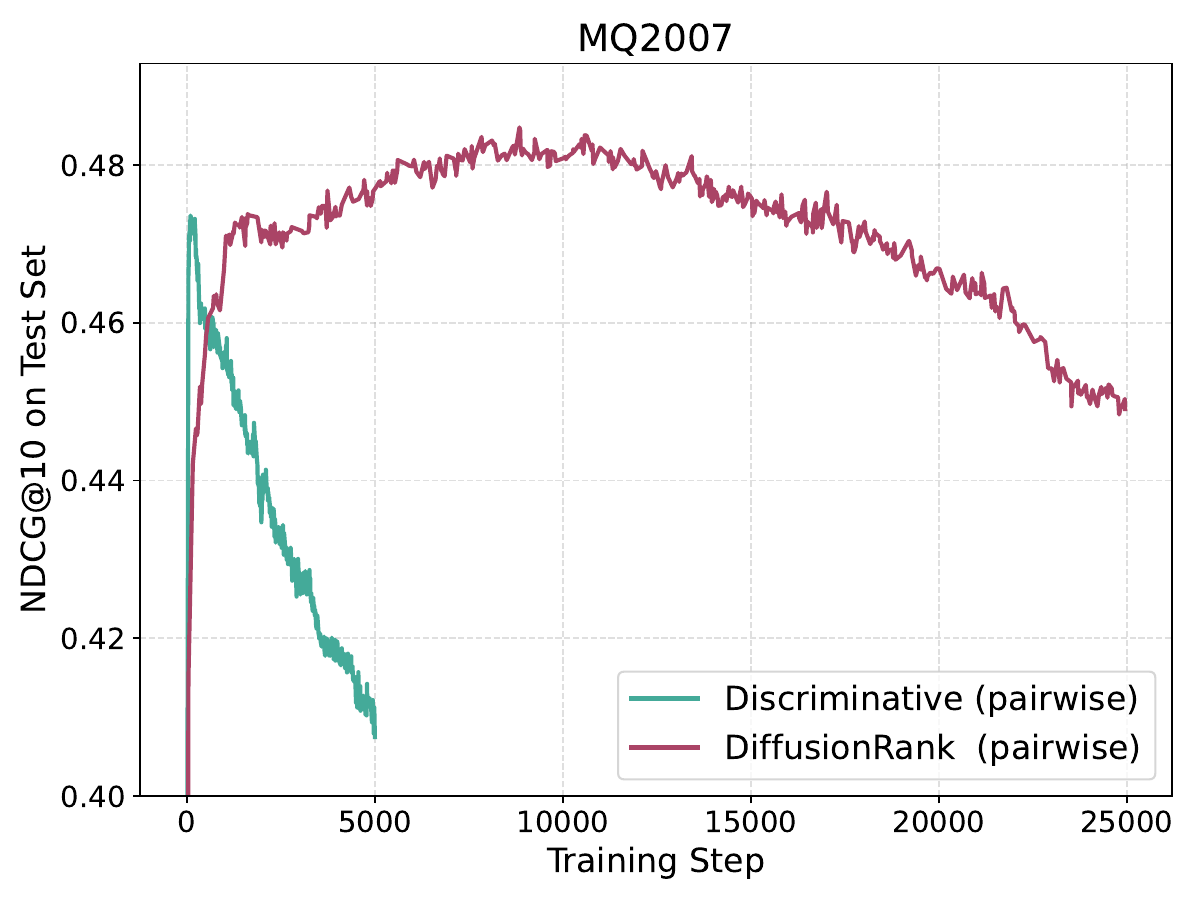}
            \caption{}
            \label{fig:training-mq2008}
        \end{subfigure}
    \end{minipage}
    \caption{Training dynamics on test data (NDCG@10) for discriminative vs.\ DiffusionRank models. We plot validation effectiveness over training steps (a) MSLR-WEB30K, (b) MSLR-WEB10K, (c) Istella-S, and (d) MQ2007. Across all datasets except Istella-S, DiffusionRank shows smoother, more stable trajectories and less pronounced degradation at later training stages, consistent with improved robustness to overfitting compared to discriminative baselines.}
    \label{fig:test_ndcg_curve}
    \vspace{-1em}
\end{figure*}

\subsection{Models}
\label{sec:experiments-model}
To ensure a principled and fair comparison between generative and discriminative approaches, we adopt the same base neural architecture across all models. This design choice allows us to isolate the effect of the learning paradigm itself without confounding factors arising from architectural differences. Both discriminative and generative models are implemented using a feedforward network (FFN) backbone with four hidden layers. The hidden layer size is set to 256 for the MQ2007 dataset and 1,024 for MSLR-WEBs, and Istella-S. Each hidden layer applies SiLU activation followed by Layer Normalization and a dropout rate of 0.1. Detailed hyperparameter tuning configurations are further elaborated in Section~\ref{sec:experiments-train}. As previously mentioned in Section~\ref{sec:model}, in DiffusionRank, the model accepts as input both noisy features and labels and predicts feature noise and the correct label.
Figure~\ref{fig:diffusionrank_models} illustrates these differences.

The XGBoost baseline~\cite{chen2016xgboost} is trained using `binary:logistic' objective for pointwise and `rank:pairwise' for pairwise tasks. Maximum tree depth is set to 6 for MQ2007 and 32 for other datasets, matching the scale of the neural models.

\subsection{Training}
\label{sec:experiments-train}
Neural models are trained using AdamW optimizer. DiffusionRank employs a continuous-time diffusion process with 50 steps for numeric and categorical features. For the categorical loss, we set $\lambda_\text{cat} = 1$ and apply an annealing scheduler on $\lambda_\text{num}$ to gradually reduce its importance during training, following TabDiff settings~\cite{shi2025tabdiff}.

The diffusion process is further parameterized with noise schedulers: `power\_mean' for numeric features and `log\_linear' for categorical features. At each step, noise is added to the features and labels, with the diffusion time variable $t$ sampled from a uniform distribution. These settings allow the model to balance feature and label denoising effectively throughout the diffusion process.

Hyperparameter tuning was performed once using random search for both the discriminative and generative models. The search space encompassed the following parameters: data normalization method (comparing the original, unnormalized data against data transformed via a Quantile Transformer), learning rate (sampled within the range of 4e-1 to 5e-8), number of hidden layers (ranging from 2 to 4), hidden layer size ({64, 128, 256, 512, 1024, 2048}), activation function (SiLU, ReLU, or Sigmoid), dropout rate ({0.0, 0.1, 0.2, 0.3, 0.4, 0.5}), and loss-weight schedules (for the generative models). Surprisingly, we observed that the majority of the optimal hyperparameters were identical across both modeling approaches, demonstrating consistent peak performance regardless of the chosen architecture. Given this consistency, we utilized the best-performing discriminative parameters for the generative models as well. Ultimately, the best model was selected based on the highest NDCG@10 score achieved on the validation set.
All models and baselines were trained on a single NVIDIA RTX A6000 GPU with 48GB of memory. Note that we built our framework on top of TabDiff's code\footnote{https://github.com/MinkaiXu/TabDiff}. To support reproducibility and facilitate future research, we made all code publicly available on GitHub\footnote{https://github.com/sadjadeb/DiffusionRank}.

\subsection{Evaluation}
\label{sec:experiments-evaluation}
We evaluate all models using standard ranking effectiveness metrics commonly adopted in Learning-to-Rank research. Specifically, we report Normalized Discounted Cumulative Gain (NDCG) and Mean Average Precision (MAP) at a cutoff of 10, with all scores averaged over queries in each dataset. While pointwise models are trained using binarized relevance labels, all evaluations are conducted using the original graded relevance judgments provided by the datasets to ensure a fair and standard assessment of ranking quality across all settings. To assess statistical significance between models, we apply a paired t-test with a significance threshold of $p < 0.05$.

\section{Results}
\label{sec:results}

\begin{table*}[]
\centering
\caption{Results of the pointwise approaches on MSLR-WEB30K, MSLR-WEB10K, Istella-S, and MQ2007. We compare discriminative baselines and DiffusionRank under pointwise training for multiple training data fractions $K$ (smaller $K$ corresponds to less training data). Performance for XGBoost is included solely for comparison with traditional models and is not part of our core hypothesis. The highest metric value for each $K$ is bolded.}

\label{tab:main_results_pointwise}
\begin{tabular}{lrrrrrrrr}
\hline
                             & \multicolumn{2}{c}{MSLR-WEB30K}                          & \multicolumn{2}{c}{MSLR-WEB10K}                          & \multicolumn{2}{c}{Istella-S}                            & \multicolumn{2}{c}{MQ2007}                               \\ \cline{2-9} 
                             & \multicolumn{1}{c}{NDCG@10} & \multicolumn{1}{c}{MAP@10} & \multicolumn{1}{c}{NDCG@10} & \multicolumn{1}{c}{MAP@10} & \multicolumn{1}{c}{NDCG@10} & \multicolumn{1}{c}{MAP@10} & \multicolumn{1}{c}{NDCG@10} & \multicolumn{1}{c}{MAP@10} \\ \hline
\textbf{K=1.0}               & \multicolumn{1}{l}{}        & \multicolumn{1}{l}{}       & \multicolumn{1}{l}{}        & \multicolumn{1}{l}{}       & \multicolumn{1}{l}{}        & \multicolumn{1}{l}{}       & \multicolumn{1}{l}{}        & \multicolumn{1}{l}{}       \\
XGBoost                      & 0.5187                      & 0.7613                     & 0.4866                      & 0.7374                     & 0.7851                      & \textbf{0.9346}            & 0.4397                      & 0.5221                     \\
Discriminative               & 0.5401                      & 0.7769                     & 0.5123                      & 0.7568                     & 0.7742                      & 0.9299                     & 0.4640                      & \textbf{0.5403}            \\
DiffusionRank                & \textbf{0.5424}             & \textbf{0.7798}            & \textbf{0.5202}             & \textbf{0.7676}            & \textbf{0.7828}             & 0.9317                     & \textbf{0.4751}             & 0.5377                     \\ \hline
\textbf{K=$\mathbf{2^{-2}}$} & \multicolumn{1}{l}{}        & \multicolumn{1}{l}{}       & \multicolumn{1}{l}{}        & \multicolumn{1}{l}{}       & \multicolumn{1}{l}{}        & \multicolumn{1}{l}{}       & \multicolumn{1}{l}{}        & \multicolumn{1}{l}{}       \\
XGBoost                      & 0.4941                      & 0.7452                     & 0.4714                      & 0.7261                     & 0.7697                      & 0.9269                     & 0.4132                      & 0.4837                     \\
Discriminative               & 0.5186                      & 0.7616                     & 0.4949                      & 0.7489                     & 0.7582                      & 0.9183                     & 0.4353                      & 0.5184                     \\
DiffusionRank                & \textbf{0.5328}             & \textbf{0.7737}            & \textbf{0.5120}             & \textbf{0.7619}            & 0.7697                      & \textbf{0.9261}            & \textbf{0.4524}             & \textbf{0.5253}            \\ \hline
\textbf{K=$\mathbf{2^{-4}}$} & \multicolumn{1}{l}{}        & \multicolumn{1}{l}{}       & \multicolumn{1}{l}{}        & \multicolumn{1}{l}{}       & \multicolumn{1}{l}{}        & \multicolumn{1}{l}{}       & \multicolumn{1}{l}{}        & \multicolumn{1}{l}{}       \\
XGBoost                      & 0.4846                      & 0.7421                     & 0.4591                      & 0.7153                     & 0.7507                      & \textbf{0.9153}            & 0.4103                      & 0.4817                     \\
Discriminative               & 0.5028                      & 0.7576                     & 0.4771                      & 0.7357                     & 0.7426                      & 0.9078                     & 0.4246                      & 0.5119                     \\
DiffusionRank                & \textbf{0.5214}             & \textbf{0.7678}            & \textbf{0.4933}             & \textbf{0.7468}            & \textbf{0.7534}             & 0.9138                     & \textbf{0.4467}             & \textbf{0.5201}            \\ \hline
\textbf{K=$\mathbf{2^{-6}}$} & \multicolumn{1}{l}{}        & \multicolumn{1}{l}{}       & \multicolumn{1}{l}{}        & \multicolumn{1}{l}{}       & \multicolumn{1}{l}{}        & \multicolumn{1}{l}{}       & \multicolumn{1}{l}{}        & \multicolumn{1}{l}{}       \\
XGBoost                      & 0.4580                      & 0.7193                     & 0.4239                      & 0.6822                     & 0.7273                      & 0.8963                     & 0.3471                      & 0.4289                     \\
Discriminative               & 0.4770                      & 0.7373                     & 0.4492                      & 0.7109                     & 0.7220                      & 0.8916                     & \textbf{0.3701}             & \textbf{0.4687}            \\
DiffusionRank                & \textbf{0.4997}             & \textbf{0.7535}            & \textbf{0.4646}             & \textbf{0.7212}            & \textbf{0.7333}             & \textbf{0.8966}            & 0.3516                      & 0.4400                     \\ \hline
\textbf{K=$\mathbf{2^{-8}}$} & \multicolumn{1}{l}{}        & \multicolumn{1}{l}{}       & \multicolumn{1}{l}{}        & \multicolumn{1}{l}{}       & \multicolumn{1}{l}{}        & \multicolumn{1}{l}{}       & \multicolumn{1}{l}{}        & \multicolumn{1}{l}{}       \\
XGBoost                      & 0.4310                      & 0.7037                     & 0.3938                      & 0.6616                     & 0.6860                      & \textbf{0.8723}            & 0.3751                      & \textbf{0.4581}            \\
Discriminative               & 0.4483                      & 0.7225                     & 0.4104                      & 0.6851                     & 0.6763                      & 0.8570                     & 0.3446                      & 0.4253                     \\
DiffusionRank                & \textbf{0.4715}             & \textbf{0.7392}            & \textbf{0.4422}             & \textbf{0.7099}            & \textbf{0.6941}             & 0.8715                     & \textbf{0.3598}             & 0.4344                     \\ \hline
\end{tabular}
\end{table*}

\begin{table*}[]
\centering
\caption{Results of the pairwise approaches on MSLR-WEB30K, MSLR-WEB10K, Istella-S, and MQ2007. We compare discriminative baselines and DiffusionRank under pairwise training for multiple training data fractions $K$ (smaller $K$ corresponds to less training data). Performance for XGBoost is included solely for comparison with traditional models and is not part of our core hypothesis. The highest metric value for each $K$ is bolded.}

\label{tab:main_results_pairwise}
\begin{tabular}{lrrrrrrrr}
\hline
                             & \multicolumn{2}{c}{MSLR-WEB30K}                          & \multicolumn{2}{c}{MSLR-WEB10K}                          & \multicolumn{2}{c}{Istella-S}                            & \multicolumn{2}{c}{MQ2007}                               \\ \cline{2-9} 
                             & \multicolumn{1}{c}{NDCG@10} & \multicolumn{1}{c}{MAP@10} & \multicolumn{1}{c}{NDCG@10} & \multicolumn{1}{c}{MAP@10} & \multicolumn{1}{c}{NDCG@10} & \multicolumn{1}{c}{MAP@10} & \multicolumn{1}{c}{NDCG@10} & \multicolumn{1}{c}{MAP@10} \\ \hline
\textbf{K=1.0}               & \multicolumn{1}{l}{}        & \multicolumn{1}{l}{}       & \multicolumn{1}{l}{}        & \multicolumn{1}{l}{}       & \multicolumn{1}{l}{}        & \multicolumn{1}{l}{}       & \multicolumn{1}{l}{}        & \multicolumn{1}{l}{}       \\
XGBoost                      & 0.4355                      & 0.6704                     & 0.4143                      & 0.6504                     & 0.5288                      & 0.7132                     & 0.4005                      & 0.4926                     \\
Discriminative               & 0.5440                      & 0.7844                     & 0.5227                      & 0.7692                     & \textbf{0.7890}             & 0.9204                     & 0.4582                      & 0.5262                     \\
DiffusionRank                & \textbf{0.5484}             & \textbf{0.7895}            & \textbf{0.5288}             & \textbf{0.7751}            & 0.7879                      & \textbf{0.9277}            & \textbf{0.4810}             & \textbf{0.5512}            \\ \hline
\textbf{K=$\mathbf{2^{-2}}$} & \multicolumn{1}{l}{}        & \multicolumn{1}{l}{}       & \multicolumn{1}{l}{}        & \multicolumn{1}{l}{}       & \multicolumn{1}{l}{}        & \multicolumn{1}{l}{}       & \multicolumn{1}{l}{}        & \multicolumn{1}{l}{}       \\
XGBoost                      & 0.4114                      & 0.6393                     & 0.4014                      & 0.6361                     & 0.5123                      & 0.6838                     & 0.4009                      & 0.4819                     \\
Discriminative               & 0.5247                      & 0.7712                     & 0.4998                      & 0.7573                     & 0.7718                      & 0.9153                     & 0.4536                      & 0.5279                     \\
DiffusionRank                & \textbf{0.5329}             & \textbf{0.7758}            & \textbf{0.5142}             & \textbf{0.7660}            & \textbf{0.7731}             & \textbf{0.9182}            & \textbf{0.4674}             & \textbf{0.5471}            \\ \hline
\textbf{K=$\mathbf{2^{-4}}$} & \multicolumn{1}{l}{}        & \multicolumn{1}{l}{}       & \multicolumn{1}{l}{}        & \multicolumn{1}{l}{}       & \multicolumn{1}{l}{}        & \multicolumn{1}{l}{}       & \multicolumn{1}{l}{}        & \multicolumn{1}{l}{}       \\
XGBoost                      & 0.4184                      & 0.6456                     & 0.4127                      & 0.6477                     & 0.5144                      & 0.7006                     & \textbf{0.4438}             & \textbf{0.5368}            \\
Discriminative               & 0.5066                      & 0.7640                     & 0.4791                      & 0.7418                     & 0.7555                      & 0.9062                     & 0.4227                      & 0.5084                     \\
DiffusionRank                & \textbf{0.5188}             & \textbf{0.7715}            & \textbf{0.4960}             & \textbf{0.7550}            & \textbf{0.7609}             & \textbf{0.9103}            & 0.4402                      & 0.5228                     \\ \hline
\textbf{K=$\mathbf{2^{-6}}$} & \multicolumn{1}{l}{}        & \multicolumn{1}{l}{}       & \multicolumn{1}{l}{}        & \multicolumn{1}{l}{}       & \multicolumn{1}{l}{}        & \multicolumn{1}{l}{}       & \multicolumn{1}{l}{}        & \multicolumn{1}{l}{}       \\
XGBoost                      & 0.3668                      & 0.6116                     & 0.3664                      & 0.5880                     & 0.5950                      & 0.7738                     & 0.3182                      & 0.3903                     \\
Discriminative               & 0.4799                      & 0.7493                     & 0.4381                      & 0.7138                     & 0.7310                      & 0.8886                     & 0.3465                      & 0.4379                     \\
DiffusionRank                & \textbf{0.4980}             & \textbf{0.7610}            & \textbf{0.4629}             & \textbf{0.7291}            & \textbf{0.7401}             & \textbf{0.8958}            & \textbf{0.3602}             & \textbf{0.4500}            \\ \hline
\textbf{K=$\mathbf{2^{-8}}$} & \multicolumn{1}{l}{}        & \multicolumn{1}{l}{}       & \multicolumn{1}{l}{}        & \multicolumn{1}{l}{}       & \multicolumn{1}{l}{}        & \multicolumn{1}{l}{}       & \multicolumn{1}{l}{}        & \multicolumn{1}{l}{}       \\
XGBoost                      & 0.3802                      & 0.6343                     & 0.3420                      & 0.5844                     & 0.5417                      & 0.7274                     & 0.3377                      & 0.4218                     \\
Discriminative               & 0.4560                      & 0.7326                     & 0.4168                      & 0.6895                     & \textbf{0.6939}             & \textbf{0.8673}            & 0.3162                      & 0.4001                     \\
DiffusionRank                & \textbf{0.4745}             & \textbf{0.7445}            & \textbf{0.4410}             & \textbf{0.7103}            & 0.6938                      & 0.8622                     & \textbf{0.3406}             & \textbf{0.4307}            \\ \hline
\end{tabular}
\end{table*}

\begin{table*}[t]
\centering
\caption{Impact of training models with perturbed data reported on MSLR-WEB30K, MSLR-WEB10K, Istella-S, and MQ2007. We compare discriminative baselines trained on clean features vs.\ perturbed features (training-time noise injection) and DiffusionRank. The best performance for each configuration is bolded. Italicized values indicate cases where training on perturbed data yielded better results than training on the original data for the discriminative baselines.}
\label{tab:noisy_features}
\resizebox{0.95\textwidth}{!}{
\begin{tabular}{lrrrrrrrr}
\hline
 & \multicolumn{2}{c}{MSLR-WEB30K}                          & \multicolumn{2}{c}{MSLR-WEB10K}                          & \multicolumn{2}{c}{Istella-S}                            & \multicolumn{2}{c}{MQ2007}                               \\ \cline{2-9} 
 & \multicolumn{1}{c}{NDCG@10} & \multicolumn{1}{c}{MAP@10} & \multicolumn{1}{c}{NDCG@10} & \multicolumn{1}{c}{MAP@10} & \multicolumn{1}{c}{NDCG@10} & \multicolumn{1}{c}{MAP@10} & \multicolumn{1}{c}{NDCG@10} & \multicolumn{1}{c}{MAP@10} \\ \hline
\textbf{Pointwise} & & & & & & & & \\
Discriminative           & 0.5400                      & 0.7769                     & 0.5123                      & 0.7568                     & 0.7742                      & 0.9299                     & 0.4640                      & 0.5403                     \\
Discriminative perturbed & 0.5395                      & 0.7765                    & \textit{0.5141}                      & \textit{0.7581}                     & \textit{0.7757}                      & 0.9289                     & 0.4539                      & 0.5212                     \\
DiffusionRank            & \textbf{0.5424}                      & \textbf{0.7798}                     & \textbf{0.5202}                      & \textbf{0.7676}                     & \textbf{0.7828}                      & \textbf{0.9317}                     & \textbf{0.4751}                      & \textbf{0.5377}                     \\
\hline
\textbf{Pairwise} & & & & & & & & \\
Discriminative            & 0.5440                      & 0.7844                     & 0.5227                      & 0.7692                     & 0.7890                      & 0.9204                     & 0.4582                      & 0.5262                     \\
Discriminative perturbed  & 0.5416                      & 0.7822                     & 0.5214                      & 0.7645                     & 0.7872                      & \textit{0.9220}                     & \textit{0.4599}                      & \textit{0.5303}                     \\
DiffusionRank             & \textbf{0.5484 }                     & \textbf{0.7895}                     & \textbf{0.5288}                      & \textbf{0.7751}                     & \textbf{0.7879}                      & \textbf{0.9277}                     & \textbf{0.4810}                      & \textbf{0.5512}                     \\ \hline
\end{tabular}}
\end{table*}

In this section, we first present the overall effectiveness of DiffusionRank compared to standard discriminative learning-to-rank baselines. We then drill down into two factors that help explain when and why generative training is beneficial: (i) the impact of training data size on DiffusionRank’s relative advantage, and (ii) an ablation that perturbs input features during discriminative training to test whether DiffusionRank’s gains could be attributed to robustness induced by noisy inputs rather than to modeling the joint feature-label distribution.

\subsection{Generative \vs Discriminative LTR}
\label{sec:results-overall}

We report the performance of different methods across all datasets in terms of NDCG@10 and MAP@10 in~\Cref{tab:main_results_pointwise} and~\Cref{tab:main_results_pairwise}. Different sections of the tables illustrate results for multiple values of $K$, reflecting different fractions of the training data used to train the model. As shown in these tables, when training with full data (i.e., $K{=}1$), DiffusionRank consistently outperforms its discriminative counterparts in both pointwise and pairwise settings on MQ2007, MSLR-WEB10K, and MSLR-WEB30K, and also on Istella-S in the pointwise setting. 
Each of these reported improvements on NDCG@10 for $K{=}1.0$ are statistically significant ($p<0.05$) according to the paired $t$-test.

Table~\ref{tab:main_results_pointwise} and~\Cref{tab:main_results_pairwise} also report performance as we progressively reduce the amount of training data. As expected, effectiveness generally degrades for all methods as $K$ decreases. This performance lead persists on MSLR-WEB10K, MSLR-WEB30K, and Istella-S, where DiffusionRank typically retains an advantage over the corresponding discriminative baselines across a wide range of data fractions, suggesting improved robustness in the low-to-moderate data regime. On the MQ2007, the trends for pairwise approaches are more mixed: DiffusionRank tends to match or improve upon discriminative baselines at moderate fractions (e.g., $K=2^{-2}$ and $K=2^{-4}$), but differences become noisier as $K$ becomes very small. Overall, these results indicate that DiffusionRank’s gains are most consistent when sufficient training data is available; in this regime, it consistently outperforms its discriminative counterparts.

To better understand these gains, Figure~\ref{fig:test_ndcg_curve} illustrates the training dynamics of discriminative and generative models. We observe a distinct trade-off in training behavior: although the generative approaches exhibit a slower learning curve, they consistently reach a higher performance ceiling in terms of NDCG@10 compared to their discriminative counterparts. While discriminative models often show widening gaps between training and testing performance, the generative models maintain closer alignment between the two. 
This increased robustness to overfitting may partially explain the consistent performance improvements observed for DiffusionRank across datasets and training paradigms.

\vspace{-1em}
\subsection{Effect of training data size}
\label{sec:results-train-size}

\begin{figure*}
    \centering
    \includegraphics[width=0.8\linewidth]{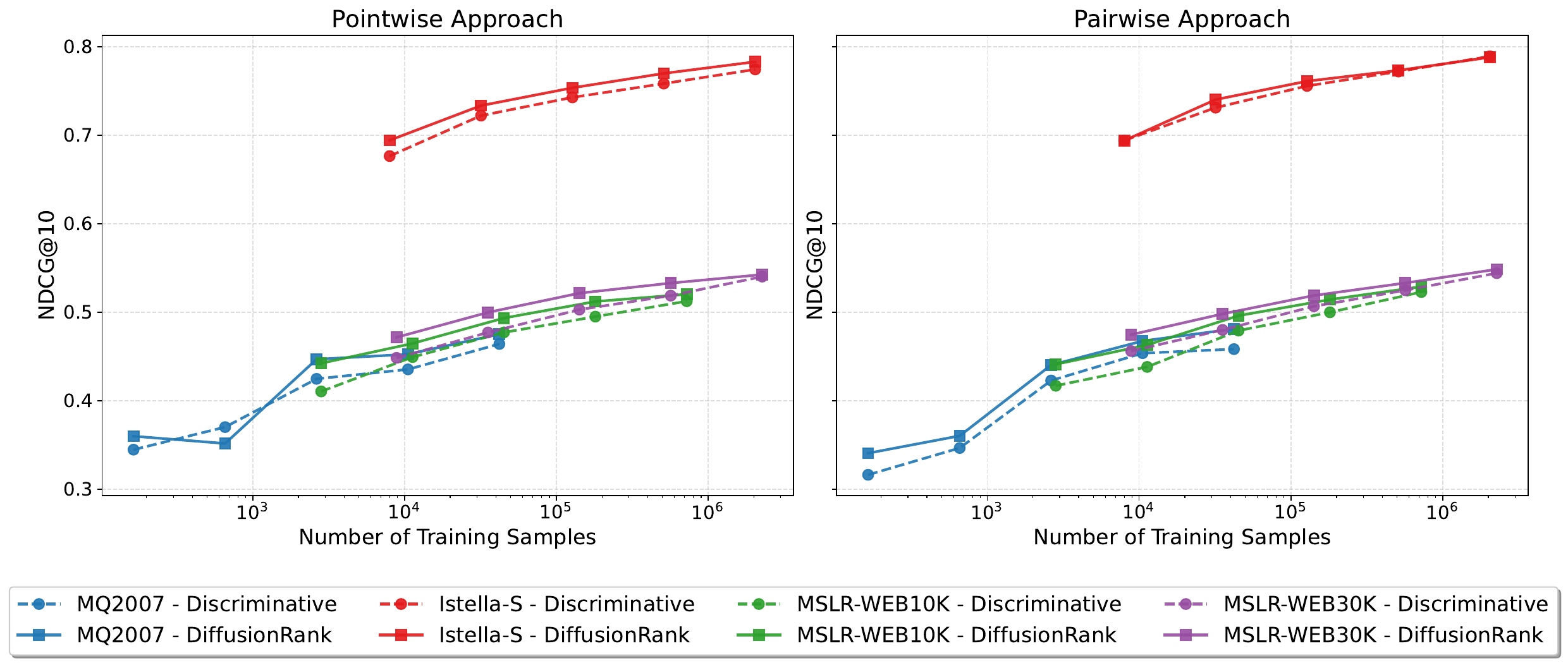}
    \caption{Effect of training set size on ranking effectiveness in terms of NDCG@10. Solid lines denote DiffusionRank and dashed lines denote the corresponding discriminative baseline.}
    \label{fig:train_size}
    \vspace{-0.5em}
\end{figure*}

We analyze the impact of training data size on ranking effectiveness in~\Cref{fig:train_size} for both pointwise and pairwise approaches. Across datasets, the relative gap between generative and discriminative training is not strictly monotonic and varies by dataset and regime, indicating that the benefits of generative training depend on both data scale and the intrinsic difficulty/noise characteristics of the dataset. On MSLR-WEB10K and MSLR-WEB30K, the separation between the solid (DiffusionRank) and dashed (discriminative) curves becomes increasingly consistent as the training set grows. This pattern suggests that DiffusionRank is able to better exploit additional supervision, potentially because modeling the joint distribution over features and relevance labels benefits from richer coverage of the feature-label space and reduces overfitting as the model sees a wider range of query-document configurations. In other words, in the larger-data regime, the generative objective appears to act as a stronger inductive bias: it encourages solutions that explain the data distribution more globally rather than fitting idiosyncrasies of limited training samples, leading to more reliable gains.

In contrast, on MQ2007, the curves exhibit noticeably higher variance, with occasional crossings in the low to moderate data regime. Moreover, in this regime, the additional modeling burden imposed by the generative objective may not always translate into improved ranking effectiveness, because there may be insufficient data to robustly estimate the joint structure that DiffusionRank is designed to capture. As a result, differences can fluctuate across training sizes, and advantages at one subsample size may not persist.

Overall, while the trends are not fully conclusive for the smaller subsets, the figure supports that DiffusionRank’s gains are most stable and pronounced when sufficient training data is available.

\subsection{Training with noisy features}
\label{sec:results-noisy}

One could posit that DiffusionRank’s improvements are a result of predicting relevance from perturbed inputs that acts as an implicit regularizer, and not because of the joint modeling of features and labels. To demonstrate that this is untrue, we train discriminative pointwise and pairwise models on artificially perturbed (noisy) features and compare them against (i) their standard, non-perturbed counterparts and (ii) DiffusionRank. \Cref{tab:noisy_features} reports NDCG@10 and MAP@10 for each dataset under these conditions. We note that in \emph{perturbed} setting, we inject controlled random noise into the input feature vector of each query--document instance during training, while keeping ground-truth relevance labels unchanged. This perturbation is applied only as a training-time augmentation, aiming to isolate whether the observed gains stem from input noise regularization rather than the generative training objective.

As shown in Table~\ref{tab:noisy_features}, introducing feature perturbations does not consistently improve discriminative learning-to-rank. In fact, discriminative models trained with noisy features perform comparably to their clean-feature counterparts. In contrast, DiffusionRank remains stronger than the discriminative models trained with noisy features, with statistically significant improvements on all four datasets: MQ2007, MSLR-WEB10K, MSLR-WEB30K, and Istella-S ($p<0.05$, paired $t$-test).
Overall, these results indicate that DiffusionRank’s gains cannot be explained solely by a regularization effect from noisy inputs. Instead, they support the conclusion that explicitly learning the joint distribution over features and relevance labels drives the effectiveness of generative LTR in our setting.

\vspace{-0.5em}
\section{Conclusion and Future Work}
\label{sec:conclusion}

In this work, we demonstrate how diffusion modeling presents an exciting new direction for generative LTR.
Our primary focus has been to empirically test the hypothesis that modeling the joint distribution of features and labels in LTR outperforms their discriminative counterparts in the pointwise and pairwise settings.
We hope this motivates future work to translate other discriminative LTR objectives, including listwise losses, to the generative setting.

Research questions for future work also include studying these approaches under large training data regimes, which may be particularly interesting as generating large volumes of training data becomes increasingly cheaper using LLMs-as-Judges~\citep{thomas2023large} and other techniques for synthetic data generation~\citep{rahmani2024synthetic, rahmani2025syndl}.
Generative LTR approaches may also be able to leverage unlabeled LTR datasets consisting only of feature vectors as part of the Gaussian diffusion process, which presents another potentially promising direction for exploration.
While our focus in this work has been on manually-designed numerical features, generative LTR may also be extended to representation-learning neural IR models where the inputs are query and document tokens.
This would require us to deal with additional modalities as part of our diffusion process; which may benefit from exploring emerging approaches for diffusion over structured datasets like DiSK~\citep{kitouni2023disk}.

While the IR community has not been immune to the excitement around emerging advancements in generative AI~\citep{white2025information}, the focus has largely been centered on the applications of LLMs, with some notable exceptions~\citep{lin2025diffusion, lin2024survey}.
We hope that work, such as ours, will uncover a broader space of opportunities to explore and apply generative modeling algorithms in IR that do not simultaneously carry the baggage of the societal risks associated with LLMs~\citep{mitra2024sociotechnical}.

\bibliographystyle{ACM-Reference-Format}
\bibliography{references}

@article{Lucchese2016PostLearningOO,
  title={Post-Learning Optimization of Tree Ensembles for Efficient Ranking},
  author={Claudio Lucchese and Franco Maria Nardini and Salvatore Orlando and R. Perego and Fabrizio Silvestri and Salvatore Trani},
  journal={Proceedings of the 39th International ACM SIGIR conference on Research and Development in Information Retrieval},
  year={2016},
  url = {https://doi.org/10.1145/2911451.2914763},
  doi = {10.1145/2911451.2914763}
}

@article{chen2016xgboost,
  title={XGBoost: A Scalable Tree Boosting System},
  author={Chen, Tianqi},
  journal={Cornell University},
  year={2016}
}

@article{qin2013introducing,
  title={Introducing LETOR 4.0 datasets},
  author={Qin, Tao and Liu, Tie-Yan},
  journal={arXiv preprint arXiv:1306.2597},
  year={2013}
}

@STRING{kdd = {Proc. SIGKDD}}

@STRING{wsdm = {Proc. WSDM}}

@STRING{sigir = {Proc. SIGIR}}

@STRING{www = {Proc. WWW}}

@STRING{cikm = {Proc. CIKM}}

@STRING{neurips = {Proc. NeurIPS}}

@STRING{acl = {Proc. ACL}}

@STRING{naacl = {Proc. NAACL}}

@STRING{icml = {Proc. ICML}}

@STRING{iclr = {Proc. ICLR}}

@STRING{emnlp = {Proc. EMNLP}}

@String{Computing = "Computing" }

@String{Computer = "{IEEE} Computer" }

@String{Springer = "Springer-Verlag" }

@ArtifactSoftware{R,
    title = {R: A Language and Environment for Statistical Computing},
    author = {{R Core Team}},
    organization = {R Foundation for Statistical Computing},
    address = {Vienna, Austria},
    year = {2019},
    url = {https://www.R-project.org/},
}

@article{robertson2009probabilistic,
  title={The probabilistic relevance framework: BM25 and beyond},
  author={Robertson, Stephen and Zaragoza, Hugo and others},
  journal={Foundations and Trends{\textregistered} in Information Retrieval},
  volume={3},
  number={4},
  pages={333--389},
  year={2009},
  publisher={Now Publishers, Inc.}
}

@inproceedings{diaz2020evaluating,
  title={Evaluating stochastic rankings with expected exposure},
  author={Diaz, Fernando and Mitra, Bhaskar and Ekstrand, Michael D and Biega, Asia J and Carterette, Ben},
  booktitle=cikm,
  pages={275--284},
  year={2020}
}

@incollection{singh:fair-pg-rank,
	Author = {Singh, Ashudeep and Joachims, Thorsten},
	Booktitle = {Advances in Neural Information Processing Systems 32},
	Editor = {H. Wallach and H. Larochelle and A. Beygelzimer and F. d\textquotesingle Alch\'{e}-Buc and E. Fox and R. Garnett},
	Pages = {5427--5437},
	Publisher = {Curran Associates, Inc.},
	Title = {Policy Learning for Fairness in Ranking},
	Year = {2019}}

@article{liu2009learning,
  title={Learning to rank for information retrieval},
  author={Liu, Tie-Yan and others},
  journal={Foundations and Trends{\textregistered} in Information Retrieval},
  volume={3},
  number={3},
  pages={225--331},
  year={2009},
  publisher={Now Publishers, Inc.}
}

@inproceedings{joachims2017unbiased,
  title={Unbiased learning-to-rank with biased feedback},
  author={Joachims, Thorsten and Swaminathan, Adith and Schnabel, Tobias},
  booktitle=wsdm,
  pages={781--789},
  year={2017}
}

@inproceedings{radlinski2008learning,
  title={Learning diverse rankings with multi-armed bandits},
  author={Radlinski, Filip and Kleinberg, Robert and Joachims, Thorsten},
  booktitle=icml,
  pages={784--791},
  year={2008}
}

@inproceedings{bruch2020stochastic,
  title={A stochastic treatment of learning to rank scoring functions},
  author={Bruch, Sebastian and Han, Shuguang and Bendersky, Michael and Najork, Marc},
  booktitle=wsdm,
  pages={61--69},
  year={2020}
}

@inproceedings{cohen2021not,
 author = {Cohen, Daniel and Mitra, Bhaskar and Lesota, Oleg and Rekabsaz, Navid and Eickhoff, Carsten},
 title = {Not All Relevance Scores are Equal: Efficient Uncertainty and Calibration Modeling for Deep Retrieval Models},
 booktitle = sigir,
 year = {2021},
  organization={ACM}
}

@article{mitra2018introduction,
  title={An introduction to neural information retrieval},
  author={Mitra, Bhaskar and Craswell, Nick},
  journal={Foundations and Trends{\textregistered} in Information Retrieval},
  year={2018},
  publisher={Now Publishers, Inc.}
}

@inproceedings{thomas2023large,
  title={Large language models can accurately predict searcher preferences},
  author={Thomas, Paul and Spielman, Seth and Craswell, Nick and Mitra, Bhaskar},
  booktitle=sigir,
  year={2024}
}

@incollection{mitra2024sociotechnical,
  title={Sociotechnical Implications of Generative Artificial Intelligence for Information Access},
  author={Mitra, Bhaskar and Cramer, Henriette and Gurevich, Olya},
  booktitle={Information Access in the Era of Generative AI},
  pages={161--200},
  year={2024},
  publisher={Springer}
}

@misc{white2025information,
  title={Information Access in the Era of Generative AI},
  author={White, Ryen W and Shah, Chirag},
  year={2025},
  publisher={Springer}
}

@article{brin1998anatomy,
  title={The anatomy of a large-scale hypertextual web search engine},
  author={Brin, Sergey and Page, Lawrence},
  journal={Computer networks and ISDN systems},
  volume={30},
  number={1-7},
  pages={107--117},
  year={1998},
  publisher={Elsevier}
}

@inproceedings{sohl2015deep,
  title={Deep unsupervised learning using nonequilibrium thermodynamics},
  author={Sohl-Dickstein, Jascha and Weiss, Eric and Maheswaranathan, Niru and Ganguli, Surya},
  booktitle={International conference on machine learning},
  pages={2256--2265},
  year={2015},
  organization={pmlr}
}

@inproceedings{yue2007support,
  title={A support vector method for optimizing average precision},
  author={Yue, Yisong and Finley, Thomas and Radlinski, Filip and Joachims, Thorsten},
  booktitle={Proceedings of the 30th annual international ACM SIGIR conference on Research and development in information retrieval},
  pages={271--278},
  year={2007}
}

@inproceedings{burges2005learning,
  title={Learning to rank using gradient descent},
  author={Burges, Chris and Shaked, Tal and Renshaw, Erin and Lazier, Ari and Deeds, Matt and Hamilton, Nicole and Hullender, Greg},
  booktitle={Proceedings of the 22nd international conference on Machine learning},
  pages={89--96},
  year={2005}
}

@article{wu2010adapting,
  title={Adapting boosting for information retrieval measures},
  author={Wu, Qiang and Burges, Christopher JC and Svore, Krysta M and Gao, Jianfeng},
  journal={Information Retrieval},
  volume={13},
  number={3},
  pages={254--270},
  year={2010},
  publisher={Springer}
}

@article{song2019generative,
  title={Generative modeling by estimating gradients of the data distribution},
  author={Song, Yang and Ermon, Stefano},
  journal={Advances in neural information processing systems},
  volume={32},
  year={2019}
}

@article{ho2020denoising,
  title={Denoising diffusion probabilistic models},
  author={Ho, Jonathan and Jain, Ajay and Abbeel, Pieter},
  journal={Advances in neural information processing systems},
  volume={33},
  pages={6840--6851},
  year={2020}
}

@article{fonseca2023tabular,
  title={Tabular and latent space synthetic data generation: a literature review},
  author={Fonseca, Joao and Bacao, Fernando},
  journal={Journal of Big Data},
  volume={10},
  number={1},
  pages={115},
  year={2023},
  publisher={Springer}
}

@article{zheng2022diffusion,
  title={Diffusion models for missing value imputation in tabular data},
  author={Zheng, Shuhan and Charoenphakdee, Nontawat},
  journal={arXiv preprint arXiv:2210.17128},
  year={2022}
}

@article{hernandez2022synthetic,
  title={Synthetic data generation for tabular health records: A systematic review},
  author={Hernandez, Mikel and Epelde, Gorka and Alberdi, Ane and Cilla, Rodrigo and Rankin, Debbie},
  journal={Neurocomputing},
  volume={493},
  pages={28--45},
  year={2022},
  publisher={Elsevier}
}

@inproceedings{assefa2020generating,
  title={Generating synthetic data in finance: opportunities, challenges and pitfalls},
  author={Assefa, Samuel A and Dervovic, Danial and Mahfouz, Mahmoud and Tillman, Robert E and Reddy, Prashant and Veloso, Manuela},
  booktitle={Proceedings of the First ACM International Conference on AI in Finance},
  pages={1--8},
  year={2020}
}

@article{borisov2022language,
  title={Language models are realistic tabular data generators},
  author={Borisov, Vadim and Se{\ss}ler, Kathrin and Leemann, Tobias and Pawelczyk, Martin and Kasneci, Gjergji},
  journal={arXiv preprint arXiv:2210.06280},
  year={2022}
}

@inproceedings{liu2023goggle,
  title={Goggle: Generative modelling for tabular data by learning relational structure},
  author={Liu, Tennison and Qian, Zhaozhi and Berrevoets, Jeroen and van der Schaar, Mihaela},
  booktitle=iclr,
  year={2023}
}

@article{xu2019modeling,
  title={Modeling tabular data using conditional gan},
  author={Xu, Lei and Skoularidou, Maria and Cuesta-Infante, Alfredo and Veeramachaneni, Kalyan},
  journal={Advances in neural information processing systems},
  volume={32},
  year={2019}
}

@inproceedings{kotelnikov2023tabddpm,
  title={Tabddpm: Modelling tabular data with diffusion models},
  author={Kotelnikov, Akim and Baranchuk, Dmitry and Rubachev, Ivan and Babenko, Artem},
  booktitle={International conference on machine learning},
  pages={17564--17579},
  year={2023},
  organization={PMLR}
}

@inproceedings{lee2023codi,
  title={Codi: Co-evolving contrastive diffusion models for mixed-type tabular synthesis},
  author={Lee, Chaejeong and Kim, Jayoung and Park, Noseong},
  booktitle={International Conference on Machine Learning},
  pages={18940--18956},
  year={2023},
  organization={PMLR}
}

@article{kitouni2023disk,
  title={Disk: A diffusion model for structured knowledge},
  author={Kitouni, Ouail and Nolte, Niklas and Hensman, James and Mitra, Bhaskar},
  journal={arXiv preprint arXiv:2312.05253},
  year={2023}
}

@inproceedings{kim2022sos,
  title={Sos: Score-based oversampling for tabular data},
  author={Kim, Jayoung and Lee, Chaejeong and Shin, Yehjin and Park, Sewon and Kim, Minjung and Park, Noseong and Cho, Jihoon},
  booktitle={Proceedings of the 28th ACM SIGKDD conference on knowledge discovery and data mining},
  pages={762--772},
  year={2022}
}

@inproceedings{zhang2024mixed,
  title={Mixed-type tabular data synthesis with score-based diffusion in latent space},
  author={Zhang, Hengrui and Zhang, Jiani and Srinivasan, Balasubramaniam and Shen, Zhengyuan and Qin, Xiao and Faloutso, Christos and Rangwala, Huzefa and Karypis, George},
  booktitle=iclr,
  year={2024}
}

@inproceedings{shi2025tabdiff,
  title={TabDiff: a Mixed-type Diffusion Model for Tabular Data Generation},
  author={Juntong Shi and Minkai Xu and Harper Hua and Hengrui Zhang and Stefano Ermon and Jure Leskovec},
  booktitle=iclr,
  year={2025}
}

@article{mcgee2012yes,
  title={Yes, Bing Has Human Search Quality Raters and Here’s How They Judge Web Pages},
  author={McGee, Matt},
  journal={Search Engine Land},
  url={http://searchengineland.com/bing-search-quality-rating-guidelines-130592},
  year={2012}
}

@article{fuhr1989optimum,
  title={Optimum polynomial retrieval functions based on the probability ranking principle},
  author={Fuhr, Norbert},
  journal={ACM Transactions on Information Systems (TOIS)},
  volume={7},
  number={3},
  pages={183--204},
  year={1989},
  publisher={ACM New York, NY, USA}
}

@inproceedings{cossock2006subset,
  title={Subset ranking using regression},
  author={Cossock, David and Zhang, Tong},
  booktitle={International conference on computational learning theory},
  pages={605--619},
  year={2006},
  organization={Springer}
}

@article{li2007mcrank,
  title={Mcrank: Learning to rank using multiple classification and gradient boosting},
  author={Li, Ping and Wu, Qiang and Burges, Christopher},
  journal={neurips},
  volume={20},
  year={2007}
}

@article{chen2009ranking,
  title={Ranking measures and loss functions in learning to rank},
  author={Chen, Wei and Liu, Tie-Yan and Lan, Yanyan and Ma, Zhi-Ming and Li, Hang},
  journal={neurips},
  volume={22},
  year={2009}
}

@article{herbrich2000large,
  title={Large margin rank boundaries for ordinal regression},
  author={Herbrich, Ralf and Graepel, Thore and Obermayer, Klaus},
  journal={Advances in Large Margin Classifiers},
  year={2000},
  publisher={The MIT Press}
}

@article{freund2003efficient,
  title={An efficient boosting algorithm for combining preferences},
  author={Freund, Yoav and Iyer, Raj and Schapire, Robert E and Singer, Yoram},
  journal={Journal of machine learning research},
  volume={4},
  number={Nov},
  pages={933--969},
  year={2003}
}

@article{li2025comprehensive,
  title={A comprehensive survey of image generation models based on deep learning},
  author={Li, Jun and Zhang, Chenyang and Zhu, Wei and Ren, Yawei},
  journal={Annals of Data Science},
  volume={12},
  number={1},
  pages={141--170},
  year={2025},
  publisher={Springer}
}

@article{wang2024challenges,
  title={Challenges and opportunities of generative models on tabular data},
  author={Wang, Alex X and Chukova, Stefanka S and Simpson, Colin R and Nguyen, Binh P},
  journal={Applied Soft Computing},
  volume={166},
  pages={112223},
  year={2024},
  publisher={Elsevier}
}

@article{minaee2024large,
  title={Large language models: A survey},
  author={Minaee, Shervin and Mikolov, Tomas and Nikzad, Narjes and Chenaghlu, Meysam and Socher, Richard and Amatriain, Xavier and Gao, Jianfeng},
  journal={arXiv preprint arXiv:2402.06196},
  year={2024}
}

@inproceedings{cui2025recent,
  title={Recent advances in speech language models: A survey},
  author={Cui, Wenqian and Yu, Dianzhi and Jiao, Xiaoqi and Meng, Ziqiao and Zhang, Guangyan and Wang, Qichao and Guo, Steven Y and King, Irwin},
  booktitle={acl},
  pages={13943--13970},
  year={2025}
}

@article{xing2024survey,
  title={A survey on video diffusion models},
  author={Xing, Zhen and Feng, Qijun and Chen, Haoran and Dai, Qi and Hu, Han and Xu, Hang and Wu, Zuxuan and Jiang, Yu-Gang},
  journal={ACM Computing Surveys},
  volume={57},
  number={2},
  pages={1--42},
  year={2024},
  publisher={ACM New York, NY}
}

@inproceedings{wang2017irgan,
  title={IRGAN: A minimax game for unifying generative and discriminative information retrieval models},
  author={Wang, Jun and Yu, Lantao and Zhang, Weinan and Gong, Yu and Xu, Yinghui and Wang, Benyou and Zhang, Peng and Zhang, Dell},
  booktitle={sigir},
  pages={515--524},
  year={2017}
}

@article{goodfellow2014generative,
  title={Generative adversarial nets},
  author={Goodfellow, Ian J and Pouget-Abadie, Jean and Mirza, Mehdi and Xu, Bing and Warde-Farley, David and Ozair, Sherjil and Courville, Aaron and Bengio, Yoshua},
  journal={neurips},
  volume={27},
  year={2014}
}

@article{yu2023depth,
  title={An in-depth study on adversarial learning-to-rank},
  author={Yu, Hai-Tao and Piryani, Rajesh and Jatowt, Adam and Inagaki, Ryo and Joho, Hideo and Kim, Kyoung-Sook},
  journal={Information Retrieval Journal},
  volume={26},
  number={1},
  pages={1},
  year={2023},
  publisher={Springer}
}

@inproceedings{park2019adversarial,
  title={Adversarial sampling and training for semi-supervised information retrieval},
  author={Park, Dae Hoon and Chang, Yi},
  booktitle={www},
  pages={1443--1453},
  year={2019}
}

@article{li2022learning,
  title={Learning to rank method combining multi-head self-attention with conditional generative adversarial nets},
  author={Li, Jinzhong and Zeng, Huan and Peng, Lei and Zhu, Jingwen and Liu, Zhihong},
  journal={Array},
  volume={15},
  pages={100205},
  year={2022},
  publisher={Elsevier}
}

@article{deshpande2020evaluating,
  title={Evaluating a generative adversarial framework for information retrieval},
  author={Deshpande, Ameet and Khapra, Mitesh M},
  journal={arXiv preprint arXiv:2010.00722},
  year={2020}
}

@incollection{jain2020improving,
  title={Improving Convergence in IRGAN with PPO},
  author={Jain, Moksh and Kamath, S Sowmya},
  booktitle={Proceedings of the 7th ACM IKDD CoDS and 25th COMAD},
  pages={328--329},
  year={2020}
}

@inproceedings{lu2019psgan,
  title={Psgan: A minimax game for personalized search with limited and noisy click data},
  author={Lu, Shuqi and Dou, Zhicheng and Jun, Xu and Nie, Jian-Yun and Wen, Ji-Rong},
  booktitle={sigir},
  pages={555--564},
  year={2019}
}

@article{yu2021diagnostic,
  title={Diagnostic evaluation of policy-gradient-based ranking},
  author={Yu, Hai-Tao and Huang, Degen and Ren, Fuji and Li, Lishuang},
  journal={Electronics},
  volume={11},
  number={1},
  pages={37},
  year={2021},
  publisher={MDPI}
}

@article{li2024listwise,
  title={Listwise learning to rank method combining approximate NDCG ranking indicator with Conditional Generative Adversarial Networks},
  author={Li, Jinzhong and Zeng, Huan and Xiao, Cunwei and Ouyang, Chunjuan and Liu, Hua},
  journal={Pattern Recognition Letters},
  volume={179},
  pages={31--37},
  year={2024},
  publisher={Elsevier}
}

@phdthesis{hofmann2013fast,
  title={Fast and reliable online learning to rank for information retrieval},
  author={Hofmann, Katja},
  year={2013},
  school={University of Amsterdam}
}

@inproceedings{zhuang2021interpretable,
  title={Interpretable ranking with generalized additive models},
  author={Zhuang, Honglei and Wang, Xuanhui and Bendersky, Michael and Grushetsky, Alexander and Wu, Yonghui and Mitrichev, Petr and Sterling, Ethan and Bell, Nathan and Ravina, Walker and Qian, Hai},
  booktitle={wsdm},
  pages={499--507},
  year={2021}
}

@inproceedings{rahmani2024synthetic,
  title={Synthetic test collections for retrieval evaluation},
  author={Rahmani, Hossein A and Craswell, Nick and Yilmaz, Emine and Mitra, Bhaskar and Campos, Daniel},
  booktitle={sigir},
  pages={2647--2651},
  year={2024}
}

@inproceedings{rahmani2025syndl,
  title={Syndl: A large-scale synthetic test collection for passage retrieval},
  author={Rahmani, Hossein A and Wang, Xi and Yilmaz, Emine and Craswell, Nick and Mitra, Bhaskar and Thomas, Paul},
  booktitle={www},
  pages={781--784},
  year={2025}
}

@inproceedings{lin2025diffusion,
  title={Diffusion Models for Recommender Systems: From Content Distribution To Content Creation},
  author={Lin, Jianghao and Cao, Yang and Yu, Yong and Zhang, Weinan},
  booktitle={kdd},
  pages={6074--6085},
  year={2025}
}

@article{lin2024survey,
  title={A Survey on Diffusion Models for Recommender Systems},
  author={Lin, Jianghao and Liu, Jiaqi and Zhu, Jiachen and Xi, Yunjia and Liu, Chengkai and Zhang, Yangtian and Yu, Yong and Zhang, Weinan},
  journal={arXiv preprint arXiv:2409.05033},
  year={2024}
}

@inproceedings{sun2023chatgpt,
  title={Is ChatGPT Good at Search? Investigating Large Language Models as Re-Ranking Agents},
  author={Sun, Weiwei and Yan, Lingyong and Ma, Xinyu and Wang, Shuaiqiang and Ren, Pengjie and Chen, Zhumin and Yin, Dawei and Ren, Zhaochun},
  booktitle={emnlp},
  pages={14918--14937},
  year={2023}
}

@inproceedings{qin2024large,
  title={Large language models are effective text rankers with pairwise ranking prompting},
  author={Qin, Zhen and Jagerman, Rolf and Hui, Kai and Zhuang, Honglei and Wu, Junru and Yan, Le and Shen, Jiaming and Liu, Tianqi and Liu, Jialu and Metzler, Donald and others},
  booktitle={naacl},
  pages={1504--1518},
  year={2024}
}

@inproceedings{zhuang2024beyond,
  title={Beyond yes and no: Improving zero-shot llm rankers via scoring fine-grained relevance labels},
  author={Zhuang, Honglei and Qin, Zhen and Hui, Kai and Wu, Junru and Yan, Le and Wang, Xuanhui and Bendersky, Michael},
  booktitle={naacl},
  pages={358--370},
  year={2024}
}

@inproceedings{gao2025llm4rerank,
  title={Llm4rerank: Llm-based auto-reranking framework for recommendations},
  author={Gao, Jingtong and Chen, Bo and Zhao, Xiangyu and Liu, Weiwen and Li, Xiangyang and Wang, Yichao and Wang, Wanyu and Guo, Huifeng and Tang, Ruiming},
  booktitle={WWW},
  pages={228--239},
  year={2025}
}

@inproceedings{wang2025realm,
  title={REALM: Recursive Relevance Modeling for LLM-based Document Re-Ranking},
  author={Wang, Pinhuan and Xia, Zhiqiu and Liao, Chunhua and Wang, Feiyi and Liu, Hang},
  booktitle={emnlp},
  pages={23875--23889},
  year={2025}
}

@inproceedings{
song2021scorebased,
title={Score-Based Generative Modeling through Stochastic Differential Equations},
author={Yang Song and Jascha Sohl-Dickstein and Diederik P Kingma and Abhishek Kumar and Stefano Ermon and Ben Poole},
booktitle={International Conference on Learning Representations},
year={2021},
url={https://openreview.net/forum?id=PxTIG12RRHS}
}

@article{yang2023diffusurvey,
  title={Diffusion models: A comprehensive survey of methods and applications},
  author={Yang, Ling and Zhang, Zhilong and Song, Yang and Hong, Shenda and Xu, Runsheng and Zhao, Yue and Zhang, Wentao and Cui, Bin and Yang, Ming-Hsuan},
  journal={ACM Computing Surveys},
  volume={56},
  number={4},
  pages={1--39},
  year={2023},
  publisher={ACM New York, NY, USA}
}

\end{document}